\documentclass[11pt]{article}
\usepackage[margin=3.0cm]{geometry}
\usepackage[utf8]{inputenc}
\usepackage{graphicx} 
\usepackage{amsmath}
\usepackage{amssymb}
\usepackage{hyperref}
\usepackage[dvipsnames]{xcolor}
\usepackage{listings}
\usepackage{float}
\usepackage{bm}
\usepackage{kotex}
\usepackage{booktabs}
\usepackage{subcaption}
\usepackage{multirow}
\usepackage{caption}
\usepackage{comment}
\usepackage[normalem]{ulem}
\usepackage[ruled,linesnumbered,vlined]{algorithm2e}
\usepackage{hyperref}
\usepackage{cleveref}

\usepackage{cleveref}

\usepackage[
  backend=bibtex,     
  style=numeric,      
  sorting=none,       
  natbib=true,        
  maxbibnames=99      
]{biblatex}
\newcommand{\fref}[1]{Fig.~\ref{#1}}

\title{\textbf{\Large Uncertainty quantification and parameter optimization of plasma etching process using heteroscedastic Gaussian process} }

\begin{document}


\begin{center}
\textbf{\Large Uncertainty quantification and parameter optimization of plasma etching process using heteroscedastic Gaussian process}

\bigskip
Yongsu Jung$^1$, Minji Kang$^{2,3}$, Muyoung Kim$^2$, Min Sup Choi$^{3}$, Hyeong-U Kim$^{2,4,5,*}$, Jaekwang Kim$^{1,*}$\\
\bigskip
\small{
\textit{
$^1$Department of Mechanical and Design Engineering, Hongik University, Sejong, Rep.of Korea\\
$^2$Semiconductor Manufacturing Research Center, Korea Institute of Machinery and Materials,Daejeon, Rep. of Korea\\
$^3$Department of Materials Science and Engineering, Chungnam National University, Daejeon, Republic of Korea\\
$^4$Nano-Mechatronics, KIMM Campus, University of Science \& Technology (UST), Daejeon, Rep. of Korea\\
$^5$School of Mechanical Engineering, Sungkyunkwan University, Suwon, Rep. of Korea
}
}

\bigskip
\textit{
$^*$Corresponding author \\
guddn418@kimm.re.kr, jk12@hongik.ac.kr
}

\end{center}

\begin{center}
\textbf{ABSTRACT }\\
\bigskip
\begin{minipage}{0.9\textwidth}

This study presents a comprehensive framework for uncertainty quantification (UQ) and design optimization of plasma etching in semiconductor manufacturing.
The framework is demonstrated using experimental measurements of etched depth collected at nine wafer locations under various plasma conditions.
A heteroscedastic Gaussian process (hetGP) surrogate model is employed to capture the complex uncertainty structure in the data, enabling distinct quantification of (a) spatial variability across the wafer and (b) process-related uncertainty arising from variations in chamber pressure, gas flow rate, and RF power.
Epistemic uncertainty due to sparse data is further quantified and incorporated into a reliability-based design optimization (RBDO) scheme.
The proposed method identifies optimal process parameters that minimize spatial variability of etch depth while maintaining reliability under both aleatory and epistemic uncertainties.
The results demonstrate that this framework effectively integrates data-driven surrogate modeling with robust optimization, enhancing predictive accuracy and process reliability.
Moreover, the proposed approach is generalizable to other semiconductor processes, such as photolithography, where performance is highly sensitive to multifaceted uncertainties.

\end{minipage}
\end{center}


\bigskip
\begin{center}
\begin{minipage}{0.9\textwidth}
\end{minipage}
\end{center}

\newpage

\section{Introduction}

Uncertainty quantification (UQ) provides a powerful framework for improving system reliability under uncertainty, with broad applications across automotive, ocean, fluid, and civil engineering~\cite{young2010_RBDOship,JKim:2019, gu2015_RBDOautomoblie, lee2024_MFHeatExchanger}. Understanding and quantifying different sources of uncertainty is at the heart of any reliable UQ framework. In particular, two fundamentally distinct forms of error must be addressed. 
Aleatoric uncertainty reflects the inherent, irreducible randomness of the physical system, while epistemic uncertainty arises from limited knowledge. The latter is inevitably introduced when surrogate models are built—whether through insufficient data (data uncertainty) or through imperfect model structures that cannot fully capture the underlying physics (model-form uncertainty) \cite{der2009_aleatoryepistemic, hullermeier2021_aleatoryepistemic}. 
Distinguishing between the two is not merely conceptual; in high-stakes manufacturing, each type of uncertainty informs different engineering decisions, making reliability and risk analyses indispensable.

The practical implementation of UQ, however, is often constrained by the prohibitive computational cost of high-fidelity simulations and physical experiments.
To circumvent this computational hurdle, a variety of surrogate models are introduced as efficient approximations, including prominent examples like Gaussian processes (GP)~\cite{zhang2015GPR_RA}, support vector machines (SVM)~\cite{roy2023SVM_review}, and neural networks (NN)~\cite{chojaczyk2015NN_review}. 
While surrogate models provide significant computational advantages, they also entail the challenge of handling the approximation error inherent to the surrogate itself. Accordingly, a number of studies have explored strategies to address this issue, for example through active learning~\cite{echard2011_AKMCS, yang2024_sysAL}. 

Among various kinds of surrogates models, 
GP regression is the most widely used methods, valued for its data efficiency and inherent ability to quantify prediction uncertainty, making it well suited for modeling systems with sparse data \citep{schulz2018GPR_tutorial}.
The predictive capability of GP is effectively harnessed in various optimization schemes, such as Bayesian optimization with the efficient global optimization (EGO) algorithm for designing heat exchangers \citep{campet2020heatexchanger} and multi-objective optimization for strain sensor design \citep{gu2024strainsensor}. 
Despite their success in many applications, standard GP models suffer from a critical limitation: the assumption of constant noise variance (homoscedasticity), which is often violated in real-world processes.
The heteroscedastic Gaussian process (hetGP) addresses this limitation by explicitly modeling input-dependent noise (aleatory uncertainty), thereby disentangling it from the model’s own lack of knowledge (epistemic uncertainty)~\citep{goldberg1997HeteroGP, kersting2007hetGP}. 
It has been successfully employed in fields such as material structure modeling~\citep{ozbayram2024getGP_Material} and equipment degradation analysis for predicting remaining useful life~\citep{liang2024hetGP_PHM}. 


Reliability-based design optimization (RBDO) provides a systematic framework for incorporating uncertainty directly into design decisions.
Unlike deterministic optimization, RBDO seeks solutions that are not only optimal but also reliable and robust, minimizing cost while satisfying target reliability requirements~\citep{hu2016RBDO_windturbine, li2025RBDO_Hinge}. Although its direct application is often hindered by the double-loop structure requiring nested reliability analyses, surrogate models may provide efficient approximations of system responses. 
In doing so, they enable practical and computationally feasible RBDO, without compromising the rigorous treatment of uncertainty~\citep{jerez2022RBDO_structural, liu2024RBDO_lithiumbattery}.
A fundamental issue shared by both hetGP and RBDO is the rigorous distinction and propagation of different sources of uncertainty within the design framework, which has recently attracted significant research attention. 
For example, Jung et al.~\cite{jung2021CBDO_GP} incorporated surrogate model uncertainty explicitly into RBDO, while Ma et al.~\cite{ma2024RDO_uncertainty} developed a probabilistic framework to systematically decouple aleatory and epistemic uncertainties. 
From another perspective, Feng et al.~\cite{feng2021RDO_modeluncertainty} proposed a robust approach based on the robust GP with a Student-t likelihood to mitigate the impact of data contamination. 
These studies demonstrate important progress in handling uncertainty in surrogate modeling and design optimization. 
Nevertheless, a unified framework that separates aleatory and epistemic uncertainties while leveraging surrogates to capture input-dependent variance remains absent.

Within this broader theoretical context, the goal of the present study is to develop an appropriate RBDO framework for semiconductor manufacturing, with a particular focus on plasma etching processes. Optimizing such processes poses significant challenges due to the high cost and time demands of physical experimentation, which has motivated a range of alternative approaches in the literature. Early efforts relied on high-fidelity simulations to predict and optimize etch profiles \cite{huang2023process}, while subsequent studies increasingly turned to data-driven methods.
Initial data-driven attempts applied evolutionary algorithms with in-situ diagnostics to infer optimal process conditions directly from real-time plasma data~\cite{al2004optimization}. 
More recent data-driven approaches have leveraged machine learning (ML) to accelerate design exploration. For example, Guo et al.~\cite{guo2024_opt} created a surrogate model for shallow trench isolation (STI) optimization by training a NN on a large dataset from 3D etching simulations, significantly reducing prediction time. Similarly, Ko et al.~\cite{ko2023plasmaDL} developed a framework using a deep learning model trained on 2D plasma data, which was then coupled with a particle swarm optimization algorithm to efficiently identify optimal process conditions.
Data-driven techniques are not only useful for optimizing process recipes but are also indispensable for real-time control during manufacturing. For instance, Kim et al.~\cite{kim2023EndpointDetection} used a Gaussian mixture model (GMM) with optical emission spectroscopy (OES) data to accurately identify the process endpoint, a crucial step for preventing over-etching and ensuring device reliability.
GP is also widely applied in the semiconductor field; for example, Wan et al.~\cite{wan2013dynamicsampling_plasmaetch} utilized a GP-based variance metric for dynamic sampling in plasma etching, while Lang et al.~\cite{lang2021GPR_sputtering} developed a GP framework to model and optimize sputtering deposition processes for a target film thickness.
While these studies demonstrate the potential of simulations, evolutionary heuristics, and machine learning in advancing plasma etching optimization, they also reveal important limitations. Most approaches either rely heavily on physics-based simulations or require large amounts of training data, and few provide a rigorous treatment of uncertainty. In particular, the distinction between aleatory variability and epistemic model uncertainty has been largely overlooked, despite its importance for reliable process optimization. This gap motivates the present study.

\begin{figure}
\centering
\includegraphics[width=0.8\textwidth]{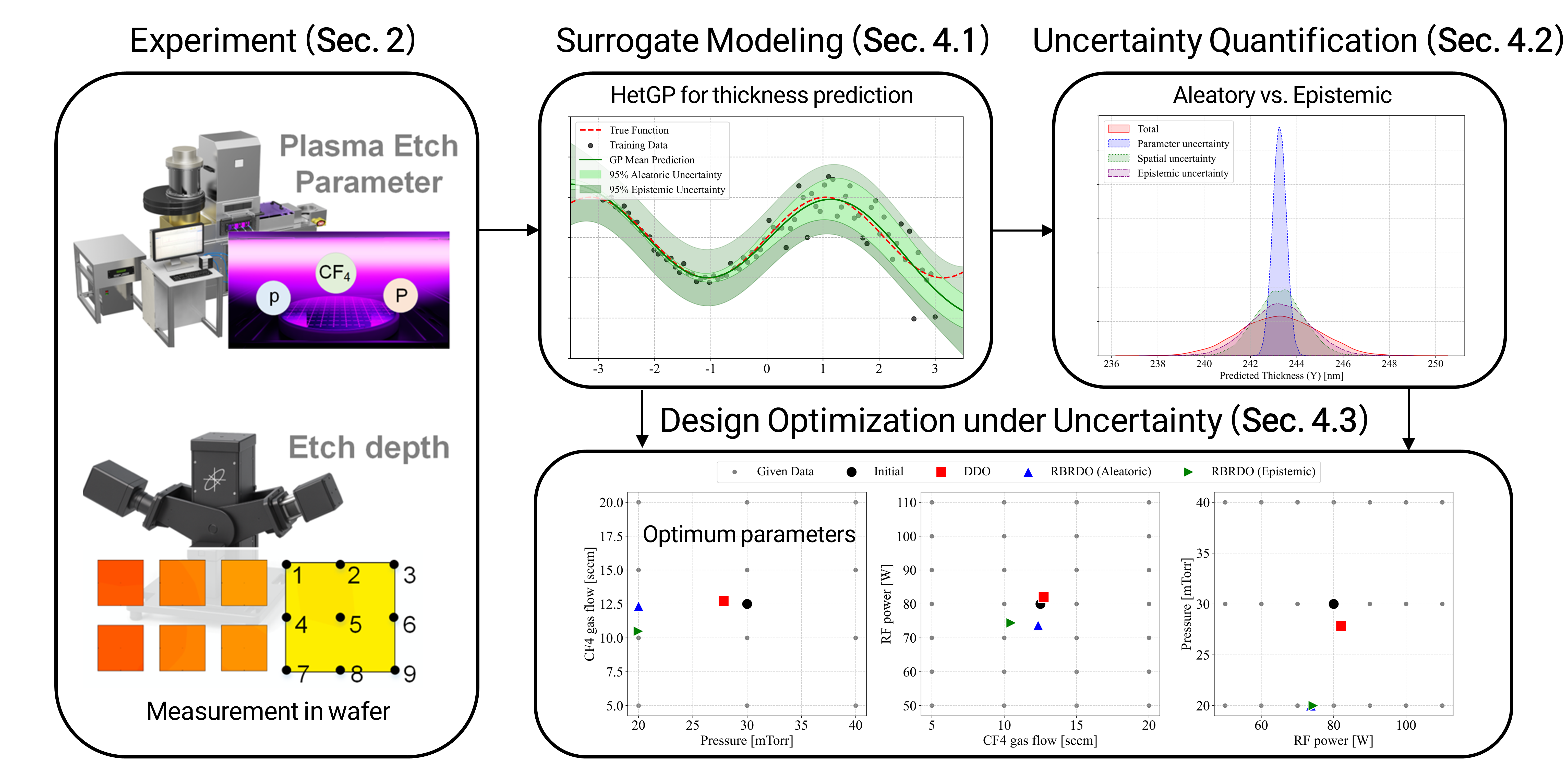}
\caption{Illustration of the hetGP-based RBDO framework.}
\label{fig:overall_framework}
\end{figure}

In this work, we develop a hetGP-based RBDO framework to maximize the spatial uniformity of etched thickness, an essential factor for device performance and yield, while satisfying a target thickness constraint. 
Although hetGP and RBDO have each been studied extensively, their integration remains underdeveloped, both theoretically and in applications such as plasma etching.
Our framework leverages high-fidelity experimental data, rather than simulations, to build hetGP surrogate models that more faithfully capture actual process behavior. These surrogate models are then used within RBDO to search for process conditions that satisfy reliability requirements.
To our knowledge, this is the first framework in plasma etching optimization that directly incorporates experimental data into RBDO while rigorously disentangling aleatory and epistemic uncertainties. 
A schematic of the overall framework proposed in this work is shown in \fref{fig:overall_framework}.
By bridging surrogate modeling and design optimization under uncertainty in this domain, the proposed framework provides a foundation for identifying and attributing the root causes of thickness variation in semiconductor manufacturing.

The remainder of this paper is organized as follows. In Section 2, the experimental setup for plasma etching and the dataset used for training and validation are introduced. Section 3 then details the proposed methodology. The results obtained from applying our framework are discussed in Section 4, where its performance is also analyzed. Finally, Section 5 concludes the paper with closing remarks and potential directions for future research.

\section{Plasma Etching Experiment and Data Generation}

The goal of this section is to provide a brief overview of the plasma etching phenomenon under investigation, the corresponding experimental setup, and the process through which the dataset was obtained.
Plasma ion etching is a technique that removes solid surfaces directionally using ions and reactive radicals generated in plasma—a state in which gas becomes electrically charged and highly reactive~\cite{Coburn2001}.
This cutting-edge technique is widely applied in the fabrication of semiconductor devices, displays, and high-aspect-ratio MEMS structures~\cite{Oehrlein2024Future,xia2022inductively, cardinaud2000plasma}. 
While we refer readers to Ref.~\citep{Donnelly2013PlasmaEtching} for a detailed historical overview, and to Ref.~\cite{Li2024PlasmaEtchingReview} for more recent development in the field, below we briefly illustrate the general physicochemical processes involved in plasma etching. 

\begin{figure}[H]
\centering
\includegraphics[width=0.7\textwidth]{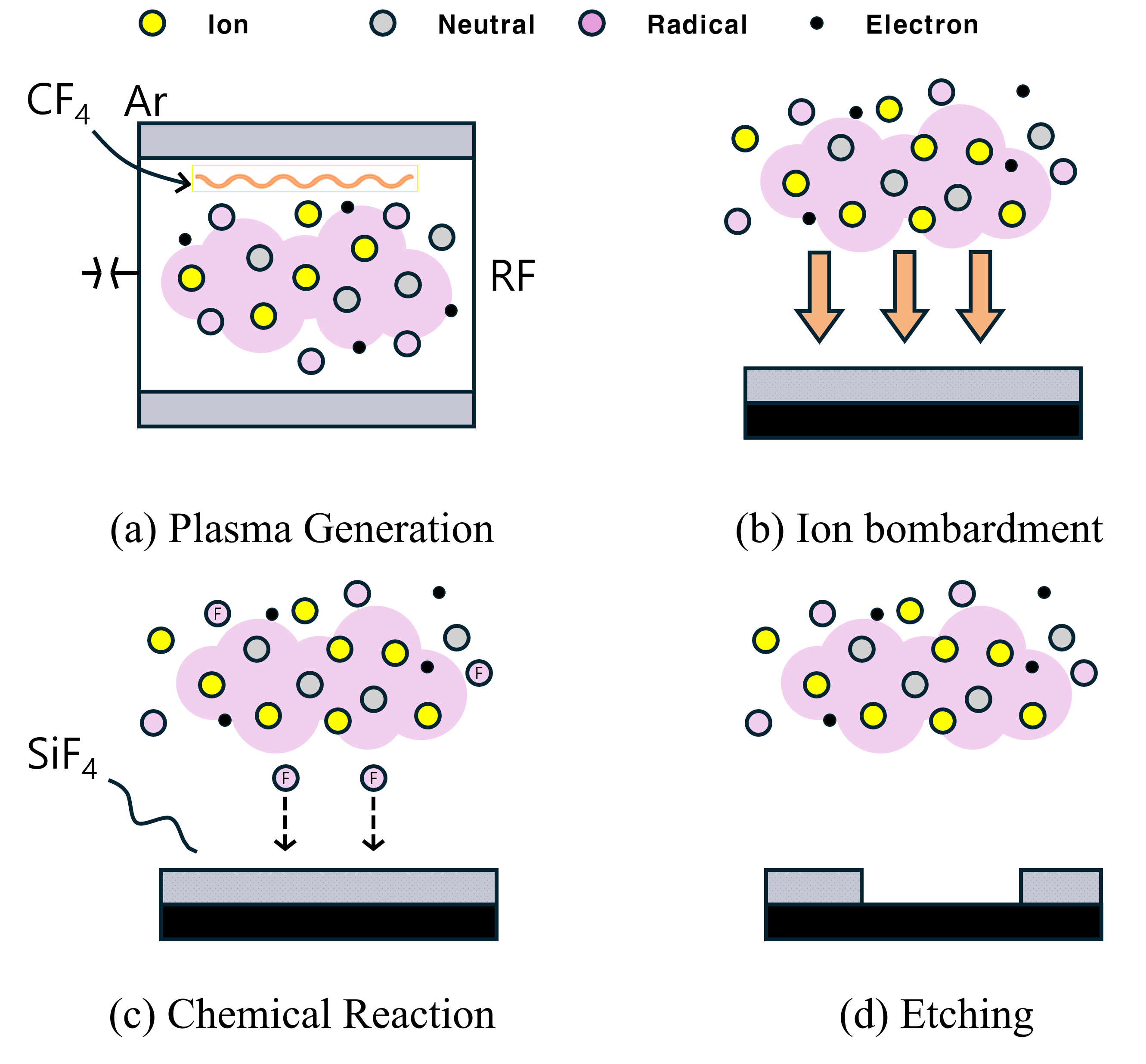}
\caption{A schematic of plasma etching process}
\label{fig:etching_process}
\end{figure}

The process begins by feeding a mixture of plasma gases into a vacuum chamber and applying a strong radio-frequency (RF) electric field; in this study, the gases considered are tetrafluoride (CF\textsubscript{4}) and argon (Ar).
Then, the RF energy knocks electrons off the gas atoms and molecules, creating a glowing cloud of ions and free electrons (the plasma). Within this plasma, CF\textsubscript{4} molecules break apart into reactive fragments (including fluorine atoms), while the argon atoms become charged but remain inert. 
This plasma environment enables etching of the solid surface through a combination of physical bombardment and chemical reactions.
Physically, the plasma’s positively charged ions are accelerated toward the surface by an electric bias, colliding with it at high speed. 
This is analogous to a sandblasting effect on a microscopic scale, as each ion impact can knock away tiny bits of the surface and help break apart its molecular bonds.
At the same time, reactive fluorine atoms from the CF\textsubscript{4} plasma swarm the surface and chemically bond with the atoms of the solid material (for example, silicon in SiO\textsubscript{2}), forming volatile compounds such as silicon tetrafluoride gas that then evaporate away
In essence, the argon provides energetic ion bombardment while the CF\textsubscript{4} provides chemically reactive fluorine, and together they enable a much faster and more effective etching process than using argon alone.
The overall mechanism of the plasma-assisted etching process is graphically illustrated in \fref{fig:etching_process}.

Although plasma etching outcomes—such as etch depth, selectivity, and anisotropy—can be empirically tuned via process parameters, quantitatively establishing physical models that link these parameters to macroscopic observables remains challenging due to scale and complexity limitations. 
This motivates the adoption of data-driven approaches for improved prediction of the etching process.
In particular, we use the experimental dataset reported in a previous work~\cite{Kang2024Etch}, which investigated plasma etching processes under varying operational conditions.

\begin{table}[h!]
  \centering
  \caption{Process conditions of ICP-RIE for $\mathrm{SiO_2}$ etching}
  \label{tab:etch_conditions}
  {
  \fontsize{10}{18}\selectfont
    \begin{tabular}{lll}
        \hline
        \textbf{Parameter} & \textbf{Unit} & \textbf{Conditions} \\ \hline
        Chamber pressure & mTorr & 20, 30, 40 \\
        Plasma power (Top) & W (watt) & 50, 60, 70, 80 , 90, 100, 110 \\
        Plasma power (Bottom) & W (watt) & 20 \\
        Process time & sec & 180 \\
        Gas $\mathrm{(CF_4)}$  & sccm & 5, 10, 15, 20 \\
        Gas (Ar) & sccm & 10 \\
        Gas $\mathrm{(O_2)}$ & sccm & 10 \\ \hline
    \end{tabular}
  }
\end{table}

For the experiments, an inductively coupled plasma (ICP) reactive ion etcher (RIE, RAINBOW 4420, Lam Research, Fremont, CA, USA) was employed. Three control variables were considered: chamber pressure ($p$ in mTorr), CF\textsubscript{4} gas flow rate ($\dot{Q}_{\mathrm{CF}_4}$ in sccm), and RF power ($P$ in W) applied to the top of the machine, which generates a high-density plasma within the RIE etcher. 
The uncertainties associated with these input parameters were estimated to be 0.15~\%, 1.0~\%, and 0.5~\%, respectively, based on the specification data provided by the equipment manufacturers.
This information will be used for the UQ and RBDO analyses, where these uncertainties are propagated through the final predictions of each etch thickness.
A 13.56 MHz radio frequency was supplied to the top copper coil to sustain the plasma discharge. Other parameters—including plasma power at the bottom electrode, process time, and gas flow rates of Ar and O\textsubscript{2}—were kept fixed. 
Optical emission spectroscopy (OES; Maya2000Pro, Ocean Insight, Orlando, FL, USA) was employed to monitor the plasma during etching. The OES system records the intensity and wavelength variations of plasma emission, and consistent OES patterns were observed, confirming plasma stability during the etching process.
Each etching run was carried out for 180 seconds under a given set of input conditions. 
The specific values of these process parameters are summarized in Table~\ref{tab:etch_conditions}.

The experimental configuration holds significant technical value, as it establishes a well-controlled plasma etching environment where key process variables—pressure, gas flow rate, and RF power—can be systematically tuned while maintaining plasma stability, 
as verified by OES diagnostics. This setup not only ensures reproducibility and reliability of the etching process but also provides a robust platform for quantitatively analyzing the sensitivity of material responses to plasma conditions. 
Furthermore, by fixing auxiliary parameters and securing stable process conditions, the setup enables consistent and systematic analysis of etching effects under different plasma conditions or with alternative sample materials in future studies.

\begin{figure}[H]
    \begin{center}
    \includegraphics[width=0.7\textwidth]{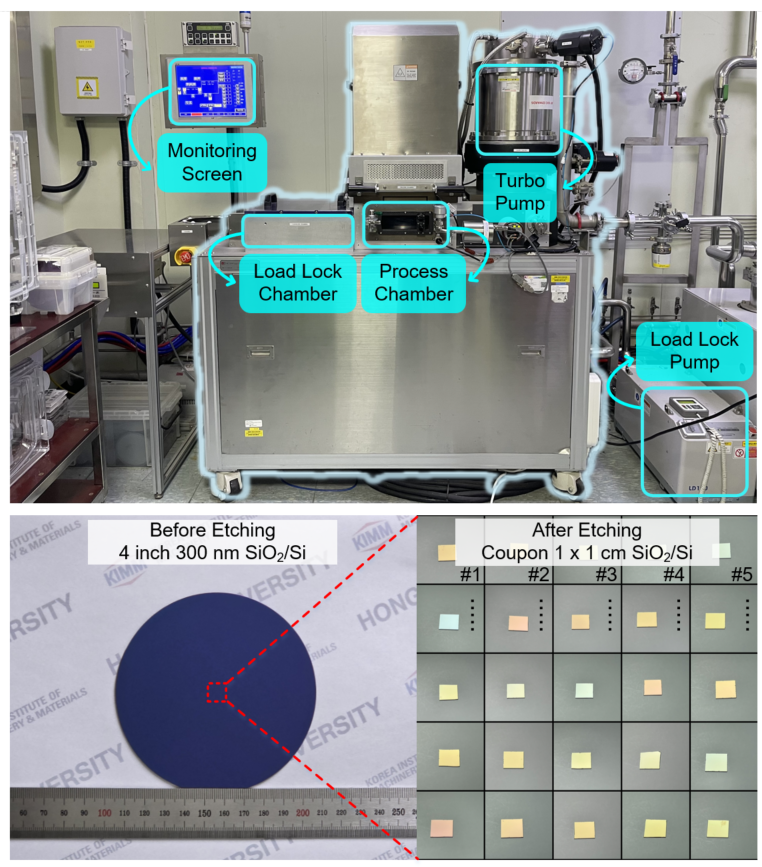}
    \end{center}
    \caption{Experimental setup of the plasma-assisted etching process and the fabricated samples.}
    \label{fig:experiments}
\end{figure}

The resulting etch depth of SiO$_2$ sample was measured at nine spatially distributed locations on the wafer using a spectroscopic ellipsometer (M-2000V, J.A. Woollam, Lincoln, Nebraska, USA).
Measurements were conducted over an energy range of 1.0–3.5 eV, which corresponds to wavelengths between 354 and 1240 nm.
This yields a nine-dimensional output vector that captures spatial variations in the remaining thickness. With the initial thickness known, the etched depth was readily calculated. 
The experimental setup and samples are shown in \Cref{fig:experiments}.
Example subsets of the dataset are shown in Table~\ref{tab:data_example}. 

\begin{figure}[H]
    \begin{center}
    \includegraphics[width=0.8\textwidth]{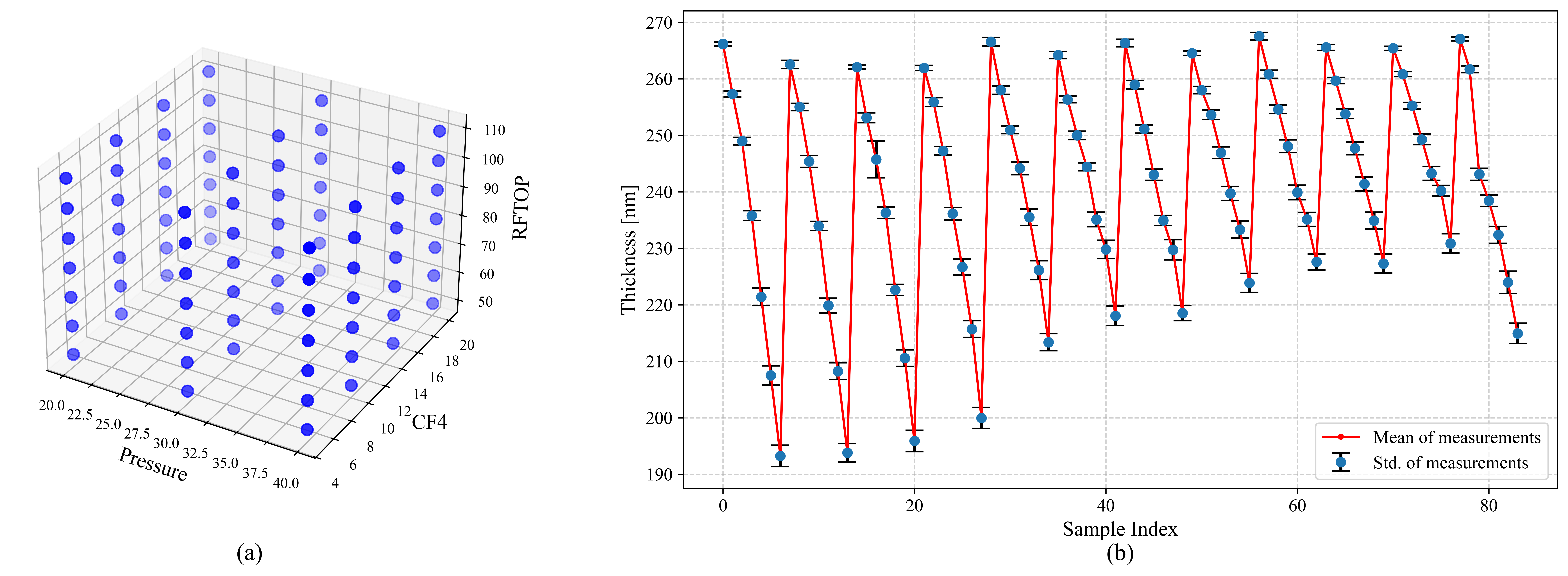}
    \end{center}
    \caption{Visualization of the training data used to construct the surrogate model. (a) Sample points in the three-dimensional process parameter space. (b) Measured thickness showing the mean and standard deviation at each sample point.}
    \label{fig:samples_in_space}
\end{figure}

\begin{figure}[H]
    \begin{center}
    \includegraphics[width=0.9\textwidth]{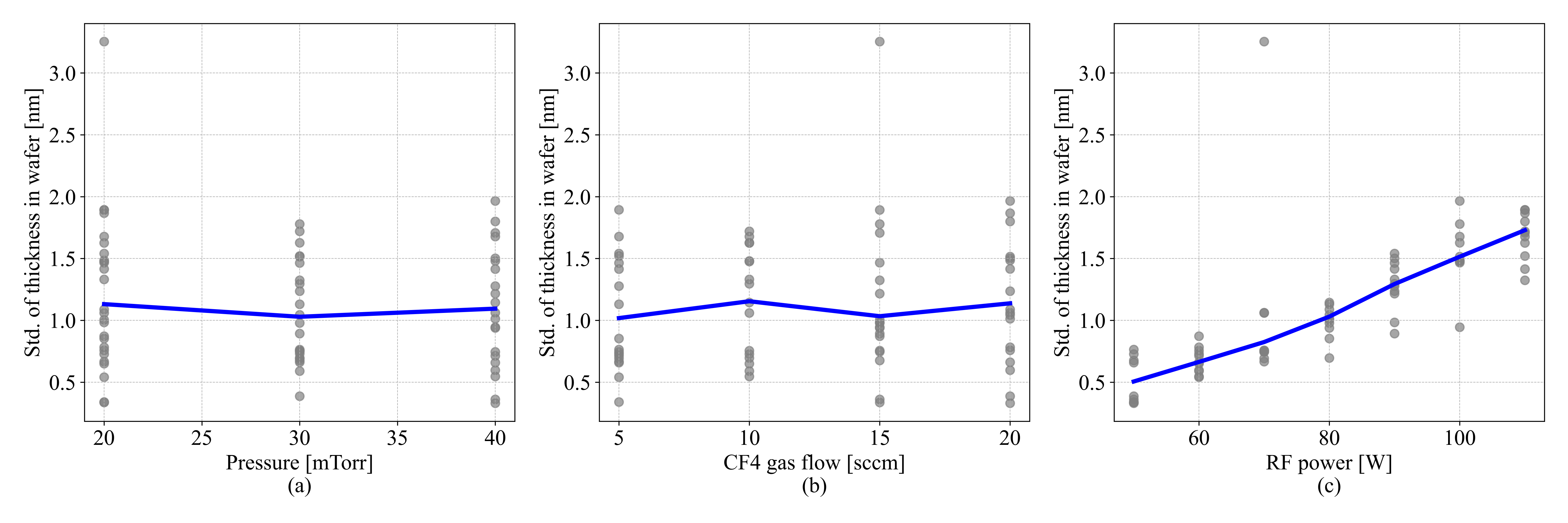}
    \end{center}
    \caption{Relationship between three input parameters and spatial variability in wafer}
    \label{fig:Noise_samples}
\end{figure}

In total, the dataset comprises 84 input–output pairs, providing the foundation for data-driven modeling of plasma-assisted etching phenomena. 
Their distribution in the process parameter space is depicted in \Cref{fig:samples_in_space}.
Note that the relative sparsity of the data points is the primary source of epistemic uncertainty, reflected as prediction variance in the hetGP surrogate model to be introduced later in this section. In contrast, aleatory uncertainty arises from spatial wafer variations and process parameter fluctuations, quantified by the standard deviation of thickness across the wafer for each experiment.
The aleatory uncertainty is analyzed in relation to the process parameters in \Cref{fig:Noise_samples}. The figure reveals a clear heteroscedastic behavior: while pressure and gas flow show weak correlation with the process variability, RF power exhibits a strong positive correlation. This direct evidence of input-dependent noise necessitates the use of a hetGP model to accurately capture both the epistemic uncertainty from sparse data and the varying aleatory uncertainty.

\begin{table}
\centering
\footnotesize
\caption{Example data set representing etched SiO$_2$ thickness}
\label{tab:data_example}
\begin{tabular}{ccc|ccccccccc}
\toprule
\multicolumn{3}{c|}{Process Parameters} & \multicolumn{9}{c}{Thickness of samples [nm]} \\
\cmidrule(r){1-3} \cmidrule(l){4-12}
$p$ [mTorr] & $\dot{Q}_{\mathrm{CF}_4}$ [sccm] & RF $P$ [W] & T\#1 & T\#2 & T\#3 & T\#4 & T\#5 & T\#6 & T\#7 & T\#8 & T\#9 \\
\midrule
20 & 5  & 50  & 266.2 & 265.6 & 266.0 & 266.2 & 266.6 & 266.7 & 266.4 & 266.0 & 265.8 \\
20 & 10 & 90  & 221.4 & 220.7 & 221.1 & 220.0 & 219.8 & 216.9 & 219.0 & 218.9 & 220.8 \\
30 & 10 & 60  & 256.8 & 256.1 & 255.9 & 255.2 & 256.0 & 256.1 & 256.9 & 257.1 & 256.9 \\
30 & 15 & 60  & 259.4 & 259.7 & 259.9 & 259.6 & 259.0 & 258.2 & 258.0 & 257.7 & 259.1\\
30 & 20 & 60  & 258.4 & 257.9 & 258.9 & 258.7 & 258.5 & 257.8 & 257.5 & 256.7 & 257.5\\
\bottomrule
\end{tabular}
\end{table}

\section{Methodology}
\label{subsec:method}

In this section, we examine the theoretical background underlying our methodology. The framework we propose rests on two fundamental pillars: GP and RBDO. To provide the necessary foundation, we review the key ideas behind each of these components. Specifically, we first introduce GP as a probabilistic surrogate modeling approach capable of capturing both mean responses and predictive uncertainties. We then discuss RBDO as a systematic optimization paradigm that incorporates reliability considerations under uncertainty. 

\subsection{Gaussian process regression}

A GP is a powerful non-parametric Bayesian method that defines a probability distribution directly over a space of latent functions. It is fully specified by a mean function $m(\mathbf{x})$ and a covariance function, or kernel function, $k(\mathbf{x}, \mathbf{x}')$, which defines the relationship between any set of points. Thus, the latent function $g(\mathbf{x})$ can be expressed as \citep{williams2006GPR}
\begin{equation}
    g(\mathbf{x}) \sim \mathcal{GP}(m(\mathbf{x}), k(\mathbf{x}, \mathbf{x}')),
\end{equation}
where $m(\mathbf{x})$ is mean function of GP. In standard GP regression, it is assumed that the response of interests is generated from the latent function $g(\mathbf{x})$ with the addition of \textit{independent and identically distributed} (I.I.D.) Gaussian noise, $\epsilon \sim \mathcal{N}(0, \sigma_n^2)$. The observation model is thus $y = g(\mathbf{x}) + \epsilon$, which defines the likelihood $p(\mathbf{y} | \mathbf{g}) = \mathcal{N}(\mathbf{y} | \mathbf{g}, \sigma_n^2 \mathbf{I})$. Given a set of $N$ training data $\mathcal{D} = \{(\mathbf{x}_i, y_i)\}_{i=1}^N$, Bayesian inference is used to form the posterior distribution over the latent function. 
This posterior, in turn, yields a predictive distribution for a new test point $\mathbf{x}_*$, which takes the form of a multivariate Gaussian distribution. 
Then, the predictive mean and variance are computed in closed form as
\begin{equation}
\mathbb{E}[g_* | \mathcal{D}] = \mathbf{k}_*^\mathrm{T} (\mathbf{K} + \sigma_n^2 \mathbf{I})^{-1} \mathbf{y},
\label{eq:gp_mean}
\end{equation}
and
\begin{equation}
\mathbb{V}[g_* | \mathcal{D}] = k(\mathbf{x}_*, \mathbf{x}_*) - \mathbf{k}_*^\mathrm{T} (\mathbf{K} + \sigma_n^2 \mathbf{I})^{-1} \mathbf{k}_*,
\label{eq:gp_variance}
\end{equation}
respectively, where $\mathbf{K} = k(\mathbf{x}, \mathbf{x})$ means the covariance matrix of the training inputs, $\mathbf{k}_* = k(\mathbf{x}, \mathbf{x}_*)$ indicates the covariance matrix between the training and test inputs, and $k(\mathbf{x}_*,\mathbf{x}_*)$ is the covariance matrix of the test inputs. 
Note that this standard formulation assumes a constant noise variance $\sigma_n^2$ across the entire input space,
a condition referred to as homoscedasticity,
which is often too restrictive for real-world applications where measurement noise can vary. 

\begin{figure}[H]
    \begin{center}
    \includegraphics[width=0.8\textwidth]{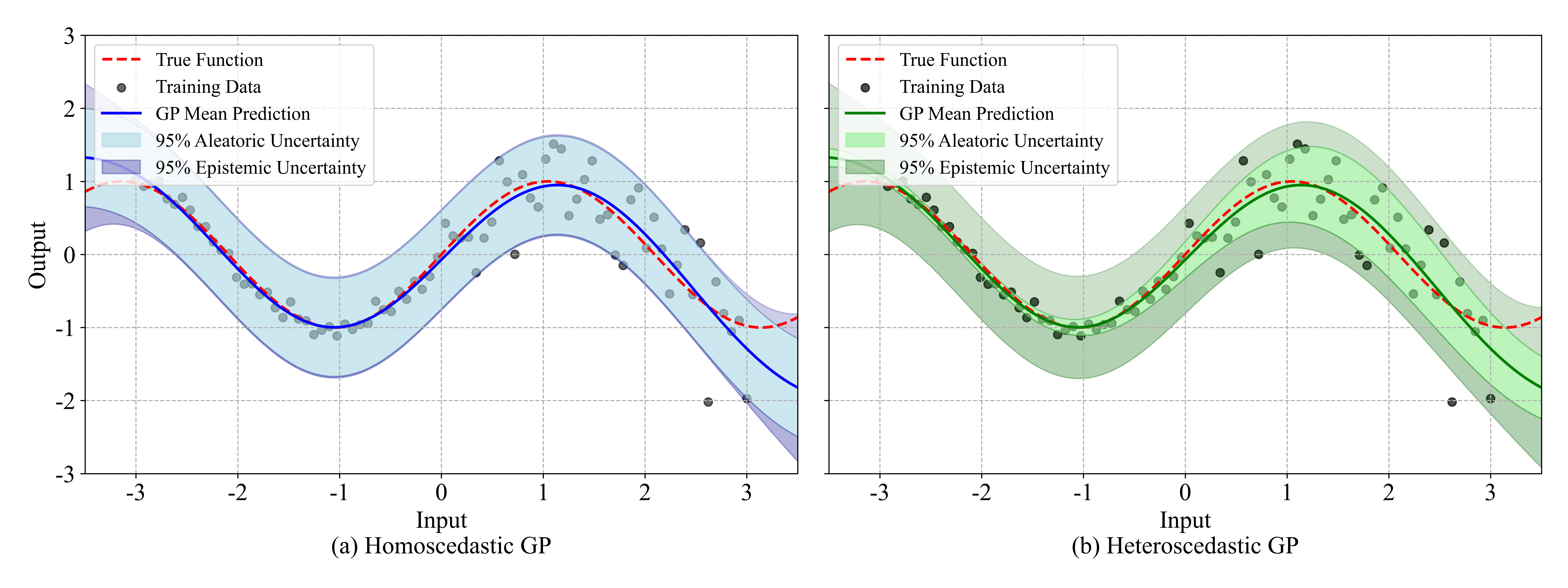}
    \end{center}
    \caption{Comparison of (a) homoscedastic GP and (b) heteroscedastic GP}
    \label{fig:homo_and_hetero_GP}
\end{figure}

To address the challenge of input-dependent noise, where the variance of observations is not constant, the hetGP has been developed \citep{goldberg1997HeteroGP, binois2018practical}. These models are generally designed to infer not only the latent mean function but also a separate function that governs the noise variance. The key difference between the two approaches is illustrated in \Cref{fig:homo_and_hetero_GP}, which contrasts a constant aleatoric uncertainty with an input-dependent one. 
For completeness, we note that stochastic Kriging also provides a framework for decomposing predictive uncertainty into distinct sources~\cite{ankenman2010stochasticKriging}, 
but our focus here remains on hetGP.

Among the various approaches, a two-stage GP framework has been widely used since it allows for the separate modeling of the mean function and the noise variance function. The underlying model for an observation $y$ at input $\mathbf{x}$ is formulated as
\begin{equation}
    y(\mathbf{x}) = g(\mathbf{x}) + \epsilon(\mathbf{x})
    \label{eq:gp_formulation}
\end{equation}
where now $\epsilon(\mathbf{x})$ is introduced as a zero-mean Gaussian noise term. 
This term variance is a function of the input expressed as
\begin{equation}
    \epsilon(\mathbf{x}) \sim \mathcal{N}(0, \sigma_n^2(\mathbf{x})), 
    \label{eq:heteroscedastic_noise}
\end{equation}
where $\mathcal{N}(\mu,\sigma^2)$ denotes the normal distribution with a mean $\mu$ and a variance $\sigma^2$.

Then, the modeling process is executed in two sequential stages. In the first stage, the mean function $g(\mathbf{x})$ by training a standard homoscedastic GP. This initial model, $g(\mathbf{x}) \sim \mathcal{GP}(m_g(\mathbf{x}), k_g(\mathbf{x}, \mathbf{x}'))$, is trained on the full dataset $\mathcal{D} = \{(\mathbf{x}_i, y_i)\}_{i=1}^N$ to obtain a posterior distribution over the latent mean function. From this the pointwise posterior mean predictions are derived as $\hat{g}(\mathbf{x}_i) = \mathbb{E}[g(\mathbf{x}_i) | \mathcal{D}]$, and the squared residuals $r_i^2 = (y_i - \hat{g}(\mathbf{x}_i))^2$ are calculated, which serve as noisy point estimates of the true underlying variance $\sigma_n^2(\mathbf{x}_i)$.

In the second stage, a new latent function $v(\mathbf{x})$ is defined as
\begin{equation}
v(\mathbf{x}) = \log(\sigma_n^2(\mathbf{x})).
\label{eq:log_variance_func}
\end{equation}
In Eq.~\eqref{eq:log_variance_func} the logarithm of the residuals is employed to ensure the non-negativity of the variance predictions.
Then, the second GP, formulated as $v(\mathbf{x}) \sim \mathcal{GP}(m_v(\mathbf{x}), k_v(\mathbf{x}, \mathbf{x}'))$, is trained to model the input-dependent noise, thereby capturing the structure of the heteroscedastic noise across the input space.

On the other hand, a more integrated approach infers the latent mean function $g(\mathbf{x})$ and the log-variance function $v(\mathbf{x})$ jointly within a single probabilistic model
~\citep{lazaro2011HeteroGP_VI}. 
In contrast to the two-stage method, the joint formulation lacks a closed-form marginal likelihood because it requires integrating over both latent GPs simultaneously. Consequently, approximate inference techniques such as variational inference (VI) are required.

In the VI framework, two separate variational distributions, which are $q(g)$ and $q(v)$, are introduced to approximate the joint posterior $p(g,v|\mathcal{D})$. The model's parameters are learned simultaneously by maximizing the evidence lower bound (ELBO) on the marginal log-likelihood given by
\begin{equation}
    \mathcal{L}(q) = \mathbb{E}_{q(g)q(v)}[\log p(\mathbf{y}|g,v)] - \text{KL}[q(g)\,\|\,p(g)] - \text{KL}[q(v)\,\|\,p(v)].
    \label{eq:elbo}
\end{equation}
It enables the model to concurrently learn the optimal hyperparameters for both the mean and variance functions, thereby ensuring that the interaction between the underlying signal and the heteroscedastic noise is properly accounted for. This, in turn, can lead to a more accurate representation of the underlying process.

The primary advantage of the two-stage approach is its simplicity and stability, as each stage involves a standard GP optimization. However, its main limitation is the decoupled nature of the estimation; errors in the initial mean estimate $\hat{g}(\mathbf{x})$ will propagate and potentially bias the subsequent estimation of the noise variance. In contrast, the joint inference approach is theoretically more principled, capturing the dependencies between the mean and variance functions. However, this comes at the cost of increased computational complexity and a non-convex optimization landscape, which can make the model more sensitive to initialization and hyperparameter settings.
Accordingly, this study adopts the two-stage approach for its simplicity, numerical stability, and computational efficiency.

Once trained, both the two stage method and VI framework use the same structure for prediction at a new test point $\mathbf{x}_*$, decomposing the total predictive variance into epistemic and aleatoric components.
The epistemic uncertainty, $\mathbb{V}_{\text{epistemic}} = \mathbb{V}[g(\mathbf{x}_*)]$, represents the model's uncertainty about the true function $g(\mathbf{x})$, while the aleatoric uncertainty, $\mathbb{V}_{\text{aleatoric}} = \exp(\mathbb{E}[v(\mathbf{x}_*)])$, represents the inherent, input-dependent noise $\sigma_n^2(\mathbf{x}_*)$. The total predictive variance is thus given by
\begin{equation}
    \mathbb{V}[y_*] = \mathbb{V}_{\text{epistemic}} + \mathbb{V}_{\text{aleatoric}}.
    \label{eq:total_variance}
\end{equation}

\subsection{Uncertainty quantification and design optimization}

In the context of UQ and reliability analysis of a system, its performance is characterized by a limit-state function, $g(\mathbf{x})$, which is dependent on a vector of random variables $\mathbf{X} = [X_1, X_2, \dots, X_n]^\mathrm{T}$, including the deterministic design variables.
These variables encapsulate sources of aleatory uncertainty, such as material properties, geometric dimensions, and external loads, and their statistical behavior is described by a joint probability density function (PDF), $f_{\mathbf{X}}(\mathbf{x})$.
Note that we are using capital letters to denote random variables, while lowercase letters represent their realizations or deterministic quantities, following the conventional notation in probability theory.
The limit-state function $g$ partitions the state space $\Omega$ into a failure region, $\Omega_f = \{\mathbf{x} \mid g(\mathbf{x}) > 0\}$ defined as the set of all outcomes $\mathbf{x}$ for which the system fails,
and a safe region $\Omega-\Omega_\mathrm{f}$.

Therefore, the probability of failure, $P_f$, which is the complement of reliability, can be estimated by integrating $f_{\mathbf{X}}(\mathbf{x})$ over the failure domain written as 
\begin{equation}
    P_f \equiv \mathrm{Pr}[G(\mathbf{X}) > 0] = \int_{G(\mathbf{X}) > 0} f_{\mathbf{X}}(\mathbf{x}) \,d\mathbf{x}.
    \label{eq:Pf}
\end{equation}
In addition, the RBDO seeks to find an optimal design that is cost-effective and reliable to aleatoric uncertainty. The formulation involves minimizing an objective function (i.e., cost function) while satisfying constraints of target reliability, which can be expressed as
\begin{equation}
    \begin{aligned}
        & \underset{\mathbf{d},\boldsymbol{\mu}_\mathbf{X}}{\text{minimize}}
        & & \text{cost}(\mathbf{d}, \boldsymbol{\mu}_\mathbf{X}) \\
        & \text{subject to}
        & & \mathrm{Pr}[G(\mathbf{d}, \mathbf{X}) > 0] \le P_{f, \text{target}}
    \end{aligned}
    \label{eq:rbdo_formulation}
\end{equation}
where $\mathbf{d}$ represents deterministic design parameters that are directly chosen by the engineer (e.g., thickness, length, cross-sectional area, or reinforcement layout), while 
$\boldsymbol{\mu}_\mathbf{X}$ reflects the controllable mean values of uncertain variables (e.g., process or material properties) that influence reliability.
Note that the conventional reliability analysis and RBDO, as presented in Eq.~\eqref{eq:Pf} and Eq.~\eqref{eq:rbdo_formulation}, primarily addresses the impact of aleatory uncertainty, which is captured by the prescribed probability distributions of the random variables $\mathbf{X}$.

However, as previously discussed,  epistemic uncertainty also inevitably arises from a lack of knowledge and data.
This second type of uncertainty includes two parts: (a) uncertainty in the PDF type and parameters; and (b) model uncertainty in the limit-state function
$G(\mathbf{X})$, when it is approximated by a surrogate model.
To account for two sources of epistemic uncertainty, 
we consider the following procedures.
Let $\boldsymbol{\theta}_X$ are the parameter vectors (such as mean and standard deviation) that determines the shape of $f_{\mathbf{X}}(\mathbf{x}; \boldsymbol{\theta}_X)$  
Since the elements of $\boldsymbol{\theta}_X$ are themselves uncertain, they are treated as random variables with an associated PDF $\pi_{\boldsymbol{\theta}}(\boldsymbol{\theta}_X)$, defined over the domain $\Omega_{\boldsymbol{\theta}}$.  
In addition, we introduce a vector $\mathbf{Z}$ to represent surrogate model uncertainty, with probability density $\pi_{\mathbf{Z}}(\mathbf{z})$ supported on $\Omega_{\mathbf{Z}}$. The surrogate limit-state function is denoted by $\hat{G}(\mathbf{x}, \mathbf{z})$, and hence the surrogate model-based failure domain is defined as $\{\hat{G}(\mathbf{x}, \mathbf{z}) > 0\}$.  
By averaging over the epistemic uncertainties in $(\boldsymbol{\theta}_X, \mathbf{Z})$, the expected probability of failure can then be expressed as  
\begin{equation}
    \mathbb{E}[P_f] = \int_{\Omega_\mathbf{Z}} \int_{\Omega_{\boldsymbol{\theta}}} \left( \int_{\hat{G}(\mathbf{x},\mathbf{z}) > 0} f_{\mathbf{X}}(\mathbf{x} | \boldsymbol{\theta}_X) \,d\mathbf{x} \right) \pi_\mathbf{Z}(\mathbf{z}) \, \pi_{\boldsymbol{\theta}}(\boldsymbol{\theta}_X) \,d\boldsymbol{\theta}_X \,d\mathbf{z}.
    \label{eq:bayesian_pf}
\end{equation}

On the other hand, the RBDO formulation that accounts for epistemic uncertainty with a target confidence level $\mathrm{CL}_{\text{target}}$ instead of using the expected probability of failure shown in Eq.~\eqref{eq:bayesian_pf} can be written as 
\begin{equation}
\begin{aligned}
    & \underset{\mathbf{d},\boldsymbol{\mu}_\mathbf{X}}{\text{minimize}}
    & & \text{cost}(\mathbf{d}, \boldsymbol{\mu}_\mathbf{X}) \\
    & \text{subject to}
    & & \int_{\Omega_\mathbf{Z}} \int_{\Omega_{\boldsymbol{\theta}}} \mathbb{I}\left[ \left( \int_{\hat{G}(\mathbf{d}, \mathbf{x}, \mathbf{z}) > 0} f_{\mathbf{X}}(\mathbf{x} | \boldsymbol{\theta}_X) \,d\mathbf{x} \right) \le P_{f, \text{target}} \right] \pi_\mathbf{Z}(\mathbf{z}) \, \pi_{\boldsymbol{\theta}}(\boldsymbol{\theta}_X) \,d\boldsymbol{\theta}_X \,d\mathbf{z} \ge \mathrm{CL}_{\text{target}}, 
\end{aligned}
\label{eq:cbdo_formulation}
\end{equation}
where $\mathbb{I}[\cdot]$ is the indicator function which returns 1 if the condition is true, otherwise 0. 
The optimization formulation in \eqref{eq:cbdo_formulation} provides a robust design framework by explicitly managing epistemic uncertainty originating from both the surrogate model and the input random variables. 
This is achieved by imposing a constraint on the confidence level.
In particular, the target probability of failure 
$P_{f, \text{target}}$ must be satisfied with a specified level of assurance.
It ensures robustness against the potential inaccuracy of the surrogate model (represented by $\mathbf{Z}$) and the parametric uncertainty of the input distributions (represented by $\boldsymbol{\theta}_X$). 
By satisfying this confidence requirement, the resulting design is less sensitive to the lack of complete knowledge, such as that arising from limited training data for the surrogate model or sparse statistical information for the inputs.

\section{Results}

This section presents the results of the proposed framework for process optimization of plasma etching, which accounts for both aleatory and epistemic uncertainties. Through the analysis, the contributions of different sources of uncertainty are quantified, and the optimal process parameters can be identified to reduce the thickness variation while satisfying the probabilistic constraint for the target thickness.

Recall that the experimental inputs are pressure, $\text{CF}_4$ gas flow rate, and RF power applied to the top of the machine, while the corresponding output is the remaining thickness of SiO\textsubscript{2} after a fixed etching time of 3 minutes.
The experimental dataset in \Cref{tab:data_example} shows that the variance is not homogeneous across the design space, indicating heteroscedastic aleatoric uncertainty. To model this effect, we employed hetGP, which extends the conventional GP by explicitly treating the noise variance as a function of the input process parameters. This enables a more faithful representation of the system behavior and a clearer separation between aleatoric and epistemic uncertainty.

It should be noted that the epistemic uncertainty associated with the input distribution parameters 
$\theta_{\mathbf{X}}$ is not accounted for, since the variability of the process parameters is prescribed by the equipment specifications.
 
\subsection{Heteroscedastic Gaussian process for thickness prediction}

The hetGP for predicting thickness is constructed using the two-stage approach, where one GP model the mean thickness, $\mu(\mathbf{x})$, and a second GP model the logarithm of its input-dependent variance, $\log\sigma^2(\mathbf{x})$, where $\mathbf{x}$ is the vector of three process parameters. Both underlying GP models employ a flexible second-degree polynomial mean function to capture global trends and a scaled radial basis function (RBF) kernel~\cite{rasmussen2006gpml} with automatic relevance determination (ARD)~\cite{wipf2007ard} to model local correlations. 
The model hyperparameters are optimized by maximizing the marginal log-likelihood using a quasi-Newton method, specifically the limited-memory Broyden–Fletcher–Goldfarb–Shanno (L-BFGS) algorithm~\cite{nocedal1980_LBFGS}. This framework enables the decomposition of the total predictive variance into its constituent parts: epistemic uncertainty, derived from the posterior variance of the mean-predicting GP, and aleatory uncertainty, obtained from the prediction of the variance-predicting GP. For a comprehensive comparison, a standard GP with an identical mean and kernel structure was developed as a baseline. The only difference is the consideration of input-dependent aleatoric uncertainty in GP modeling. All GP models are implemented and trained by GPytorch \cite{gardner2018_gpytorch}.
The experimental data was divided into a training set (85\%) and a testing set (15\%) to evaluate and compare the predictive performance of the models.

\begin{figure}[H]
    \begin{center}
    \includegraphics[width=0.8\textwidth]{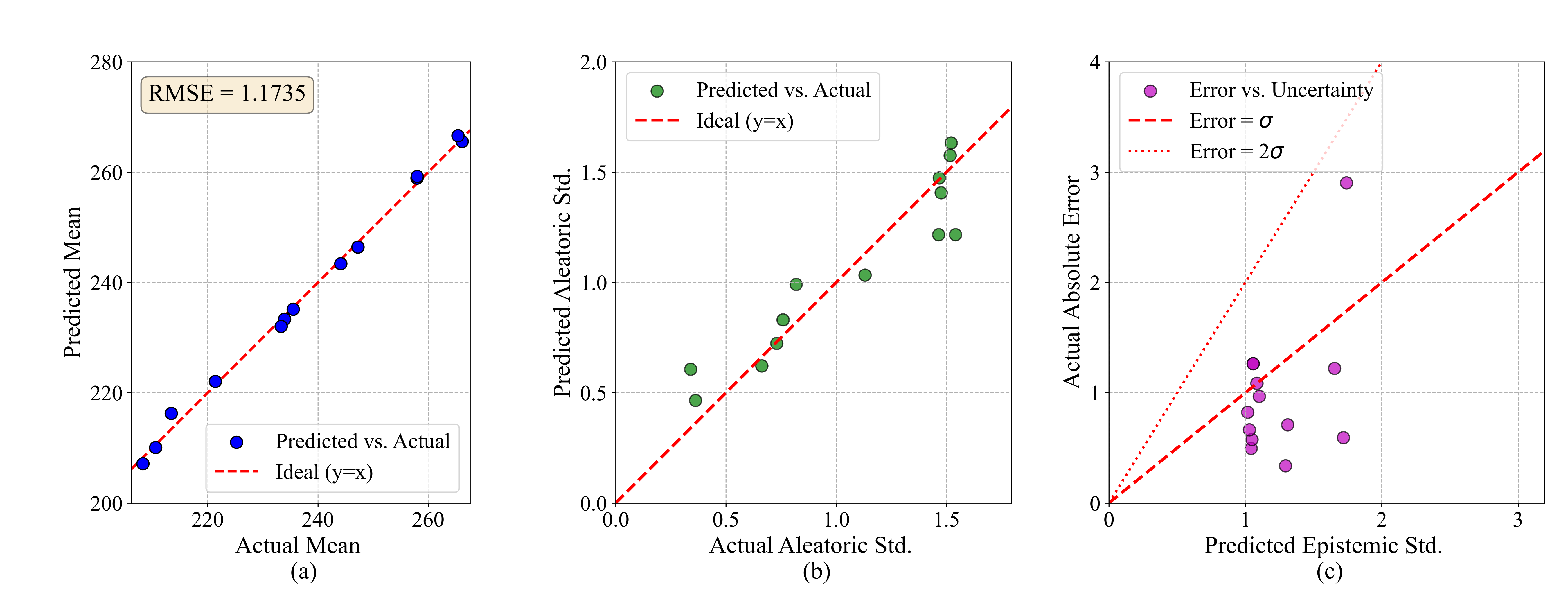}
    \end{center}
    \caption{Validation of the trained hetGP model: (a) predicted vs. actual mean thickness, (b) predicted vs. actual aleatoric standard deviation, and (c) actual absolute error vs. predicted epistemic standard deviation.}
    \label{fig:heteroGP_performance}
\end{figure}

The performance of the trained hetGP model is comprehensively validated in \Cref{fig:heteroGP_performance}.
First, \Cref{fig:heteroGP_performance}(a) compares the predicted mean thickness with the actual experimental values, showing excellent agreement along the ideal line and a low root mean square error (RMSE). More importantly, \Cref{fig:heteroGP_performance}(b) demonstrates the model's ability to capture the heteroscedastic nature of the process; the predicted aleatoric standard deviation closely matches the actual values, indicating that the model successfully learns the input-dependent noise structure. Finally, \Cref{fig:heteroGP_performance}(c) assesses the quantification of epistemic uncertainty. The plot of actual absolute error against the predicted epistemic standard deviation shows that all error points are well-bounded by the model's uncertainty estimates (specifically, within two standard deviations, $2\sigma$). This confirms that the model's confidence intervals are reliable, accurately reflecting its own predictive uncertainty.

\begin{figure}[H]
    \begin{center}
    \includegraphics[width=0.7\textwidth]{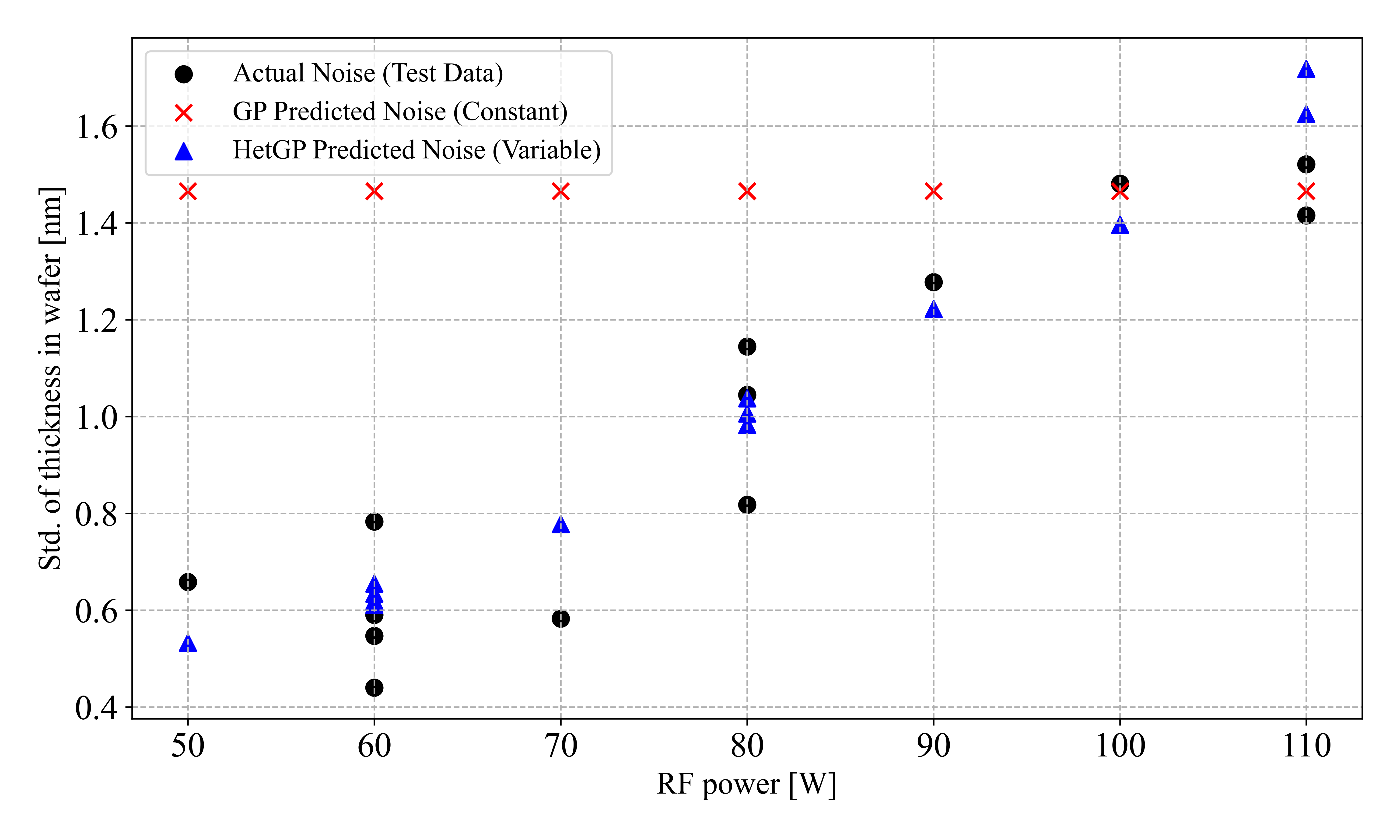}
    \end{center}
    \caption{Prediction of spatial variability in wafer using standard GP and hetGP.}
    \label{fig:heteroGP_noise}
\end{figure}

The results illustrated in \Cref{fig:heteroGP_noise} highlight a significant disparity in the performance of the two approaches for modeling aleatory uncertainty (i.e., spatial variation of thickness in wafer).
Here, plasma power is shown on the horizontal-axis and the predicted standard deviation of thickness on the vertical-axis, while the effects of chamber pressure and $\text{CF}_4$ flow rate are implicitly projected onto this 2D space. Consequently, for a fixed plasma power, multiple values of the predicted standard deviation appear, reflecting variations arising from different combinations of the other process parameters.
While the actual spatial variability in the test data indicated by black circles is clearly not constant but rather heteroscedastic, 
the standard GP only shows the constant standard deviation of the thickness in the wafer indicated by red crosses. In contrast, the hetGP framework effectively captures the input-dependent variation of the predicted thickness uncertainty, thereby fulfilling the motivation for introducing it in this study.

To provide a more rigorous and objective assessment of the standard GP and hetGP, their predictive performance was evaluated over 10 trials with different random data splits (85\% for training, 15\% for testing). The averaged performance metrics, summarized in \Cref{tab:performance_comparison}, offer a comprehensive quantitative comparison between the two surrogate models. 
First, the RMSE (Mean) for both models is identical due to the two-stage approach for hetGP, indicating that both approaches are equally proficient at predicting the mean thickness. However, this metric only evaluates point-prediction accuracy and fails to capture the models' ability to quantify uncertainty. The critical distinction emerges when evaluating the prediction of process variability. Next, the RMSE (Variance), which measures the error in predicting the input-dependent aleatoric variance, shows a stark difference: the hetGP achieves a very low error, whereas the standard GP has an error that is over 40 times larger compared to the results of the hetGP. This result decisively confirms that the assumption of constant noise is invalid for this physical process, while the hetGP successfully captures the complex, input-dependent nature of the spatial uncertainty in the wafer.

\begin{table}[b]
\centering
\caption{Performance comparison of surrogate models, averaged over 10 random data splits. The results are presented as mean $\pm$ standard deviation.}
\label{tab:performance_comparison}
\begin{tabular}{@{}lcc@{}}
\toprule
\textbf{Metric} & \textbf{hetGP} & \textbf{Standard GP} \\ \midrule
RMSE (Mean) & $1.5749 \pm 0.3249$ & $1.5749 \pm 0.3249$ \\
RMSE (Variance) & $0.4005 \pm 0.1457$ & $17.1884 \pm 4.1281$ \\
NLL (Total) & $2.0921 \pm 0.1879$ & $2.1145 \pm 0.2538$ \\
NLL (Conditional) & $1.3843 \pm 0.1239$ & $2.4019 \pm 0.0941$ \\
CRPS & $1.1150 \pm 0.1558$ & $1.3410 \pm 0.1018$ \\ \bottomrule
\end{tabular}
\end{table}

This finding is further corroborated by probabilistic scoring rules such as negative log-likelihood (NLL) and continuous ranked probability score (CRPS), which assess the quality of the entire predictive distribution \cite{gneiting2007_CRPSREF}. The hetGP demonstrates superior performance with a lower (better) CRPS compared to the standard GP. Furthermore, the NLL (Conditional), which specifically evaluates the likelihood of the variance model, is significantly lower for the hetGP, providing strong evidence for its more accurate aleatoric uncertainty quantification. It is noted that the NLL (Total) scores are closely matched despite the clear difference in variance modeling. This is because the total predictive variance, a key component in the NLL calculation, is currently dominated by the large epistemic uncertainty that arises from the insufficient number of data. As such epistemic component is significant and comparable in both models, the accurate modeling of the smaller aleatoric component by the hetGP has a less pronounced effect on the final total score. Furthermore, the NLL metric is also highly sensitive to the accuracy of the predictive mean. Given that both models achieve identical performance in mean prediction, this major source of prediction error contributes equally to both models, further explaining the similarity in their total NLL values.

In summary, the quantitative results—especially those assessing variance prediction and probabilistic accuracy—demonstrate hetGP’s ability to capture both the process mean and input-dependent variability, 
thereby establishing its robustness. 
This confirms its suitability as a surrogate model for subsequent analysis.

\subsection{Reliability analysis for thickness under aleatory and epistemic uncertainties}

To assess the reliability defined as the probability that the remaining thickness lies within a predefined range, we performed a comprehensive uncertainty quantification analysis using the trained hetGP model at an initial point, which is the center of the design space. The objective is to decompose the total predictive uncertainty into its fundamental components, which is crucial for identifying the dominant sources of variability and utilizing it in subsequent reliability analysis and design optimization. 
Building upon the general formulation discussed in Section~\ref{subsec:method}, we illustrate how the UQ and reliability analyses apply to this specific problem.

The predicted system output $\hat{y}(\mathbf{x},\mathbf{z})$ is expressed as the sum of a specific GP realization $\hat{g}(\mathbf{x},\mathbf{z})$ and the spatial variability in the wafer $\epsilon(\mathbf{x})$,
\begin{equation}
\hat{y}(\mathbf{x},\mathbf{z}) = \hat{g}(\mathbf{x},\mathbf{z}) + \epsilon(\mathbf{x}).
\label{eq:performance_model}
\end{equation}

In this work, we considered three primary sources of uncertainty as tabulated in \Cref{tab:uncertainty_class}. The first is parameter uncertainty, which represents the inherent variability of the process parameters $\mathbf{X}$ written as $f_\mathbf{X}(\mathbf{x};\boldsymbol{\theta_\mathbf{X}})$. 
Specifically, the input process parameters $\mathbf{X} = [\text{Pressure, CF}_4 \text{ gas flow, RF power}]$ are treated as independent random variables, each following a normal distribution with standard deviations of 0.15, 1.0, and 0.5~\%, respectively, as previously noted from manufacturer specifications.

The second uncertainty is spatial variability $\epsilon(\mathbf{x})$, which captures the intrinsic spatial randomness in wafer or measurement error. This type of uncertainty is modeled as lumped input-dependent aleatory uncertainty; only $\sigma^2_{\text{spatial}}(\mathbf{x})$ is used as distribution parameter in $f_\mathbf{\epsilon}(\mathbf{\epsilon};\boldsymbol{\theta_\mathbf{\epsilon}})$. Thus, it is quantified as heteroscedastic uncertainty shown in Eq.~\eqref{eq:log_variance_func}. 
The final source of uncertainty is epistemic, arising from the limited training data of the surrogate model $\hat{g}(\mathbf{x},\mathbf{z})$, expressed as $\mathbf{Z}$. In this work, it is represented by the predicted covariance given in Eq.~\eqref{eq:gp_variance}.

\begin{table}[b]
\centering
\footnotesize
\caption{Classification of uncertainty}
\label{tab:uncertainty_class}
\begin{tabular}{lll}
\toprule
\textbf{Type} & \textbf{Source} & \textbf{Modeling} \\
\midrule
\multirow{2}{*}{Aleatoric} & Process parameter uncertainty & Parametric PDF (e.g., $\mathcal{N}(\mu, \sigma^2)$) \\
\cmidrule(l){2-3}
& Spatial variability in wafer & Input-dependent model from hetGP (e.g.,  $\sigma_{spatial}^2(\mathbf{x})$) \\
\midrule
Epistemic & Surrogate model uncertainty & Predictive covariance of GP (e.g., $\Sigma_{\mathbf{x}^*}$) \\
\bottomrule
\end{tabular}
\end{table}

To quantify the contribution of each uncertainty source, we apply the law of total variance to the predictive output of the trained hetGP model,  
\begin{equation}
    \mathbb{V}(Y) = \mathbb{E}[\mathbb{V}(Y|Z)] + \mathbb{V}(\mathbb{E}[Y|Z]),
    \label{eq:total_var}
\end{equation}
where $Z$ represents a specific model realization from the GP posterior. 
In Eq.~\eqref{eq:total_var}, the first term, $\mathbb{E}[\mathbb{V}(Y|Z)]$, represents the total aleatory uncertainty (i.e., the average variance across all possible models), while the second term, $\mathbb{V}(\mathbb{E}[Y|Z])$, corresponds to the epistemic uncertainty (i.e., the variance in the mean prediction due to surrogate model uncertainty). Furthermore, the total aleatory uncertainty can be decomposed into its two constituent sources: parameter uncertainty, which arises from the variability of the input process parameters denoted as $\mathbf{X}$, and input-dependent noise, which captures the intrinsic spatial variability on the wafer denoted as $\epsilon(\mathbf{x})$. This hierarchical decomposition, summarized in \Cref{tab:variance_decomposition}, shows the attribution of variability and provides the guideline to improve reliability. For instance, if the parameter uncertainty is dominant, tightening the control of the process parameters is the most effective strategy such as a change of specification. In contrast, if the epistemic uncertainty is high, acquiring more experimental data in specific regions of the design space would be necessary to improve model confidence. Thus, it can be extended to resource allocation to determine how to maximize the information on confidence of reliability under limited budget.

\begin{table}[b]
    \footnotesize
    \centering
    \caption{Quantitative decomposition of predictive variance}
    \label{tab:variance_decomposition}
    \begin{tabular}{l r r}
        \toprule
        \textbf{Source} & \textbf{Variance [nm$^{2}$]} & \textbf{Contribution [\%]} \\
        \midrule
        \textbf{Total predictive variance} & \textbf{2.85} & \textbf{100.0} \\
        \midrule
        \multicolumn{3}{l}{\textbf{Level 1: Epistemic vs. Aleatoric}} \\
        \quad Epistemic Uncertainty & 1.70 & 59.7 \\
        \quad Aleatoric Uncertainty & 1.15 & 40.3 \\
        \midrule
        \multicolumn{3}{l}{\textbf{Level 2: In-depth aleatoric}} \\
        \quad (a) Due to Input Variability & 0.09 & 7.5 \\
        \quad (b) Due to Process Noise & 1.06 & 92.5 \\
        \bottomrule
    \end{tabular}
\end{table}

\begin{figure}[H]
    \centering
    \begin{subfigure}[b]{0.49\textwidth}
        \centering
        \includegraphics[width=\textwidth]{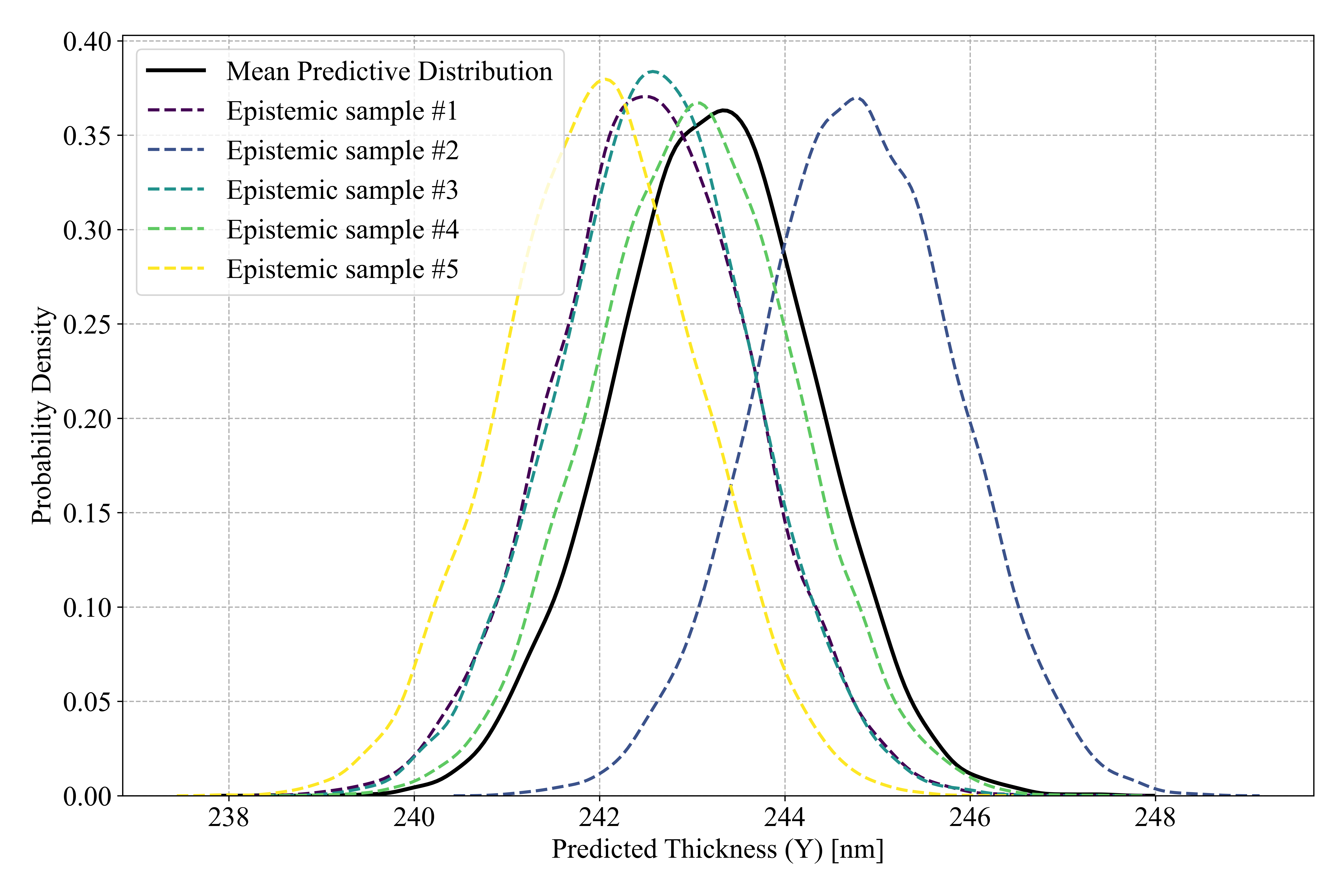}
        \subcaption{}
    \end{subfigure}
    \hfill
    \begin{subfigure}[b]{0.49\textwidth}
        \centering
        \includegraphics[width=\textwidth]{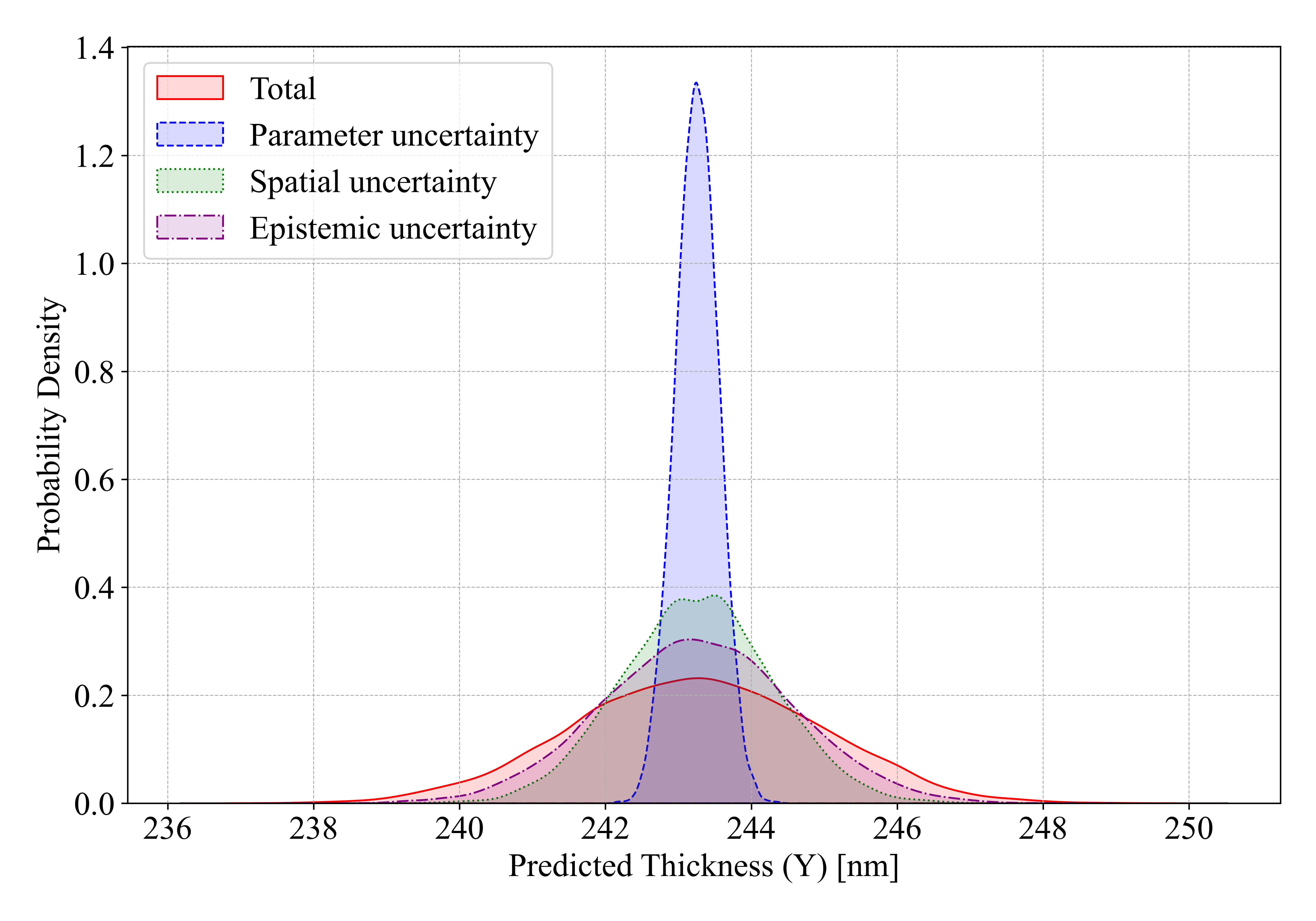}
        \subcaption{}
    \end{subfigure}
    \caption{(a) Different PDFs obtained accounting for surrogate model uncertainty and (b) Decomposition of predictive uncertainty at the initial design.}
    \label{fig:dist_full}
\end{figure}

The realization of epistemic uncertainty and the decomposition of uncertainty are demonstrated in \Cref{fig:dist_full}. 
First, \Cref{fig:dist_full}(a) illustrates five distinct predictive distributions generated by sampling five different predictions from the hetGP at an initial design point.
The resulting variation across these distributions represents the surrogate model uncertainty evaluated at the initial process parameters.
On the other hand, \Cref{fig:dist_full}(b) visualizes how each decomposed uncertainty source affects the predictive distribution of thickness. Each distribution is plotted while the other uncertainty sources are held fixed. The distribution accounting for epistemic uncertainty exhibits the largest variation. This observation agrees with the variance decomposition results shown in \Cref{tab:variance_decomposition}. 
It again confirms that the epistemic uncertainty, stemming from the sparse dataset, is the dominant contributor to the total predictive variance at this design point.

Now, reliability analysis accounting for aleatory and epistemic uncertainties can be performed. To this end, we define the conditional probability of failure, $P_f(\mathbf{Z})$, for the GP model realization (i.e., given $\mathbf{z}$ shown in \eqref{eq:cbdo_formulation}). This probability is due to aleatoric uncertainty only which can be expressed as
\begin{equation}
    \begin{aligned}
        P_f(\mathbf{z}) 
        &= \mathrm{Pr}\left[ | (\hat{G}(\mathbf{X},\mathbf{z}) + \epsilon) - T_{\text{target}} | > \Delta \right] \\
        &= \int \int \mathbb{I}\left[|(\hat{G}(\mathbf{x}, \mathbf{z}) + \epsilon) - T_{\text{target}}| > \Delta\right] \cdot f_{\mathbf{X}}(\mathbf{x}) \cdot f_{\epsilon}(\epsilon | \mathbf{x}) \,d\mathbf{x} \,d\epsilon
        \label{eq:conditional_pf}
    \end{aligned}
\end{equation}
where $\mathbb{I}[\cdot]$ is the indicator function, and $\Delta$ means a maximum tolerance. The distribution of $P_f(\mathbf{z})$ represents the impact of epistemic uncertainty on the reliability assessment, and its PDF of the probability of failure can be expressed as
\begin{equation}
f_{P_f}(p) = \int \delta\left(p - P_f(\mathbf{z})\right) f_{\mathbf{Z}}(\mathbf{z}|\mathcal{D}) \,d\mathbf{z}.
\label{eq:pf_pdf_dirac}
\end{equation}
Crucially, this formulation reveals that the probability of failure is not a single value but a random variable, $P_f(\mathbf{z})$, whose distribution is induced by the epistemic uncertainty $\mathbf{Z}$ of the surrogate model. While a confidence-based framework could utilize this distribution directly to satisfy a probabilistic constraint, such an approach significantly increases computational complexity and causes the ambiguity in the criterion for setting the target confidence. Therefore, for practical application within an optimization loop, this study adopts a more direct approach by marginalizing the distribution into a single, representative metric which is the augmented probability of failure, $\bar{P}_f$. The uncertainty is marginalized over the distribution of $\mathbf{X}$ and $\mathbf{Z}$ written as
\begin{equation}
    \begin{aligned}        
    \bar{P}_{f} &\equiv \mathbb{E}_{\mathbf{X,Z},\epsilon}[|(\hat{G}(\mathbf{x,z}) + \epsilon) - T_{\text{target}}| > \Delta] \\
                &= \int\int_{\Omega_f} f_{\mathbf{X}}(\mathbf{x} | \boldsymbol{\theta}_X) \, f_{\mathbf{Z}}(\mathbf{z}|\mathcal{D}) f_\epsilon(\epsilon|\mathbf{x}) \,d\mathbf{x}d\mathbf{z}d\epsilon
    \label{eq:total_pf}
    \end{aligned}
\end{equation}
and effectively averages out the epistemic uncertainty and serves as a robust, single-point estimate for subsequent design optimization. It is noted that surrogate model uncertainty is dependent on $\mathbf{x}$, and thus double-loop calculation is inevitable even though augmented probability of failure is employed.

\subsection{Design optimization of process parameters under aleatory and epistemic uncertainties}

In this section, the proposed formulation for reliability-based robust design optimization (RBRDO) of the plasma etching process accounting for epistemic uncertainty is introduced to reduce the variance of remaining thickness (i.e., robustness) while satisfying the probabilistic constraints of the target thickness range (i.e., reliability).
To demonstrate the proposed methodology, an example problem is set with a thickness of 240 nm, a tolerance of 2.4 nm and a failure probability of 3.0~\%. 
The initial design point is defined as $\mathbf{x}^0 = [\text{Pressure, CF}_4 \text{ gas flow, RF power}] = [30.0\; \text{mTorr}, 12.5\; \text{sccm}, 80.0 \; \text{W]}$, which was estimated from the experimental results by simple visual inspection.
For brevity, the units of the variables are omitted hereafter.

To begin with, it is instructive to consider a conventional deterministic design optimization (DDO) as a baseline. It disregards all sources of uncertainty and aims to find a set of process parameters that drives the model's predicted output to the exact target value. Using the mean prediction of the trained surrogate model, $\hat{g}(\mu_\mathbf{X})$, the DDO problem is formulated as
\begin{equation}
    \begin{aligned}
        & \underset{\boldsymbol{\mu}_\mathbf{X}}{\text{minimize}}
        & & (\hat{y}(\boldsymbol{\mu}_\mathbf{X}) - T_{\text{target}})^2
    \end{aligned}
    \label{eq:DDO_formulation}
\end{equation}
where $\mu_\mathbf{X}$ represents the design vector.
Note that the design bounds are omitted in the optimization formulation for brevity.
While computationally efficient, the DDO approach provides an optimum that is only valid under nominal conditions. It offers no guarantee of performance or reliability when the inherent aleatory and epistemic uncertainties are considered, often leading to designs that are sensitive to small variations and may fail in practice. 

In comparison, the proposed RBRDO for the plasma etching process accounting for three types of uncertainty explained in previous section can be formulated as
\begin{equation}
    \begin{aligned}
        & \underset{\mu_\mathbf{X}}{\text{minimize}}
        & & \mathbb{V}\left[\hat{y}(\mathbf{X},\mathbf{\bar{z}})\right] \\
        & \text{subject to}
        & & \mathrm{Pr}\left[ |\hat{y}(\mathbf{X},\mathbf{Z}) - T_{\text{target}}| > \Delta \right] \le P_{f, \text{target}}
    \end{aligned}
    \label{eq:RBRDO_PROCES}
\end{equation}
where the outer and inner probabilistic measures indicate epistemic uncertainty and aleatory uncertainty, respectively. In the above formulation, $\hat{y}(\mathbf{X},\mathbf{Z})$ is the surrogate model prediction for the thickness shown in Eq.~\eqref{eq:performance_model}. The objective is to minimize the variance of the thickness, thereby enhancing the process robustness by reducing its sensitivity to the variation of parameters. Note that the variance of thickness is only affected by aleatory uncertainty. To solve the proposed RBRDO problem in \eqref{eq:RBRDO_PROCES}, a computational framework is established based on the trained two-stage hetGP model explained in Section 4.1. The nested probability constraint, which is the core of the formulation, is evaluated using a double-loop Monte Carlo simulation (MCS). The outer loop quantifies the epistemic uncertainty by sampling multiple realizations of the hetGP model,  or each of these model realizations. The inner loop then propagates the aleatory uncertainty of the input parameters $\mathbf{X}$ to estimate the probability of failure. Detailed algorithm can be found in \Cref{alg:rbrdo}.

The score function method allows for the efficient estimation of the gradient of an expectation. The gradient of the probability of failure $P_f$ with respect to the mean of the input parameters $\boldsymbol{\mu_X}$ can be derived as~\cite{lee2011_RBDO} 
\begin{equation}
    \begin{aligned}
        \nabla_{\mu_\mathbf{X}} P_f &= \int \mathbb{I}(\hat{y}(\mathbf{x},\mathbf{z}) > T_{\text{target}}) \nabla_{\mu_\mathbf{X}} f_{\mathbf{X}}(\mathbf{x}) \,d\mathbf{x} \\
        &= \int \mathbb{I}(\hat{y}(\mathbf{x},\mathbf{z}) > T_{\text{target}}) f_{\mathbf{X}}(\mathbf{x}) \nabla_{\mu_\mathbf{X}} \log f_{\mathbf{X}}(\mathbf{x}) \,d\mathbf{x} \\
        &= \mathbb{E}_{\mathbf{X}} \left[ \mathbb{I}(\hat{y}(\mathbf{x},\mathbf{z}) > T_{\text{target}}) \nabla_{\mu_\mathbf{X}} \log f_{\mathbf{X}}(\mathbf{X}) \right].
    \end{aligned}
    \label{eq:score_function}
\end{equation}
For a normal distribution $\mathbf{X} \sim \mathcal{N}(\boldsymbol{\mu_X}, \Sigma_{\mathbf{X}})$, the score function is $\nabla_{\boldsymbol{\mu_X}} \log f_{\mathbf{X}} = \Sigma_{\mathbf{X}}^{-1}(\mathbf{x}-\boldsymbol{\mu_X})$. This results in a practical MCS for the gradient, which can be computed in a single simulation run alongside the value of $P_f$ itself. Detailed descriptions on sampling-based RBDO and its score function can be found in Ref.~\cite{lee2011_RBDO}.

\begin{algorithm}[H]
 \small
 \caption{RBRDO accounting for aleatory and epistemic uncertainties}
 \label{alg:rbrdo}
 \SetAlgoLined
 
 \KwIn{
    A GP surrogate model $\hat{y}(\cdot)$, Initial design mean $\bm{\mu}_{\mathbf{X}}^{(0)}$, Sample sizes $N_{\text{ale}}, N_{\text{epi}}$ for aleatory and epistemic loops, Target thickness $T_{\text{target}}$, Tolerance of target thickness $\Delta$, Convergence-tolerance $\epsilon$\;
 }
 \KwOut{Final optimum design $\bm{\mu}_{\mathbf{X}, \text{epi}}^*$}
 \BlankLine

 \tcp{Stage 1: RBRDO with aleatory uncertainty (RBRDO-A)}
 Initialize $k \leftarrow 0$, $converge \leftarrow \text{False}$\;
 \While{\textup{not} converge}{
    Draw $N_{\text{ale}}$ samples: $\mathbf{x}^{(i)} \sim \mathcal{N}(\bm{\mu}_{\mathbf{X}}^{(k)}, \mathbf{\Sigma}_{\mathbf{X}})$ for $i=1, \dots, N_{\text{ale}}$\;   
    Generate responses: $t_i \gets \mathbb{E}[\hat{y}(\mathbf{x}^{(i)})]$ for $i=1, \dots, N_{\text{ale}}$ \tcp*{GP mean prediction}
    $J^{(k)} \gets \mathbb{V}[\{t_i\}_{i=1}^{N_{\text{ale}}}]$ \tcp*{Objective} 
    $P_f^{(k)} \gets \frac{1}{N_{\text{ale}}}\sum_{i=1}^{N_{\text{ale}}} \mathbb{I}(|t_i - T_{\text{target}}| > \Delta)$ \tcp*{Constraint} 
    Compute $\nabla J^{(k)}$ and $\nabla P_f^{(k)}$\;
    $\bm{\mu}_{\mathbf{X}}^{(k+1)} \gets \text{UpdateDesign}(\bm{\mu}_{\mathbf{X}}^{(k)}, J^{(k)}, P_f^{(k)}, \nabla J^{(k)}, \nabla P_f^{(k)})$\;
    \If{$||\bm{\mu}_{\mathbf{X}}^{(k+1)} - \bm{\mu}_{\mathbf{X}}^{(k)}|| \le \epsilon$}{
       $converge \leftarrow \text{True}$\;
    }
    $k \leftarrow k + 1$\;
 }
 $\bm{\mu}_{\mathbf{X}, \text{ale}}^* \gets \bm{\mu}_{\mathbf{X}}^{(k)}$\;
 \BlankLine
 
 \tcp{Stage 2: RBRDO with aleatory and epistemic uncertainties (RBRDO-AE)}
 Initialize $k \leftarrow 0$, $converge \leftarrow \text{False}$, and $\bm{\mu}_{\mathbf{X}}^{(0)} \gets \bm{\mu}_{\mathbf{X}, \text{ale}}^*$\;
 \While{\textup{not} converge}{
    Draw $N_{\text{epi}}$ GP realizations: $z^{(j)} \sim \mathcal{N}(0,1)$ for $j=1, \dots, N_{\text{epi}}$ \tcp*{Outer loop}
    Draw $N_{\text{ale}}$ input samples: $\mathbf{x}^{(i)} \sim \mathcal{N}(\bm{\mu}_{\mathbf{X}}^{(k)}, \mathbf{\Sigma}_{\mathbf{X}})$ for $i=1, \dots, N_{\text{ale}}$ \tcp*{Inner loop}
    \For{$i \leftarrow 1$ \KwTo $N_{\text{ale}}$}{
        \For{$j \leftarrow 1$ \KwTo $N_{\text{epi}}$}{
            $t_{ij} \gets \hat{y}_j(\mathbf{x}^{(i)},\mathbf{z}^{(j})$ \tcp*{Response for i-th input from j-th GP realization}
        }
    }
    
    Generate mean responses: $\bar{t}_i \gets \frac{1}{N_{\text{epi}}}\sum_{j=1}^{N_{\text{epi}}} t_{ij}$ for $i=1, \dots, N_{\text{ale}}$\;
    $J_{\text{epi}}^{(k)} \gets \mathbb{V}[\{\bar{t}_i\}_{i=1}^{N_{\text{ale}}}]$ \tcp*{Objective} 
    $P_{f, \text{epi}}^{(k)} \gets \frac{1}{N_{\text{ale}} N_{\text{epi}}} \sum_{i=1}^{N_{\text{ale}}} \sum_{j=1}^{N_{\text{epi}}} \mathbb{I}(|t_{ij} - T_{\text{target}}| > \Delta)$ \tcp*{Constraint} 
    Compute $\nabla J_{\text{epi}}^{(k)}$ and $\nabla P_{f, \text{epi}}^{(k)}$\;
    $\bm{\mu}_{\mathbf{X}}^{(k+1)} \gets \text{UpdateDesign}(\bm{\mu}_{\mathbf{X}}^{(k)}, J_{\text{epi}}^{(k)}, P_{f, \text{epi}}^{(k)}, \nabla J_{\text{epi}}^{(k)}, \nabla P_{f, \text{epi}}^{(k)})$\;
    \If{$||\bm{\mu}_{\mathbf{X}}^{(k+1)} - \bm{\mu}_{\mathbf{X}}^{(k)}|| \le \epsilon$}{
       $converge \leftarrow \text{True}$\;
    }
    $k \leftarrow k + 1$\;
 }
 $\bm{\mu}_{\mathbf{X}, \text{epi}}^* \gets \bm{\mu}_{\mathbf{X}}^{(k)}$\;
 \BlankLine
 \Return{$\bm{\mu}_{\mathbf{X}, \text{epi}}^*$}\;
\end{algorithm}

Now, the results of RBRDO for process parameter are summarized in \Cref{tab:optimization_summary}, where target tolerance and probability of failure are set to 2.4~nm and 3.0~\%, respectively. 
Here, we compare four design approaches: the initial design, DDO, RBRDO considering only aleatory uncertainty (RBRDO-A), and the proposed RBRDO considering both aleatory and epistemic uncertainties (RBRDO-AE). 
As expected, the naive choice of the initial design point is highly unreliable ($P_{f}=75.35~\%$), indicating that it is unlikely to satisfy the remaining thickness requirement.
The DDO finds a feasible solution but with a relatively high process variance. The key trade-off is evident when comparing the two RBRDO methods. The RBRDO-A approach, by neglecting model uncertainty, identifies an optimum with the lowest cost (the objectctive function of the present optimization, i.e., variance of 0.9136) that pushes to the absolute limit of the reliability constraint ($P_{f}=2.94\%$). In contrast, the proposed RBRDO-AE accounts for the surrogate model uncertainty. It yields a more robust design that comfortably satisfies the reliability target ($P_{f}=0.71\%$) by accepting a slightly higher cost (variance of 0.9626). 
This result highlights the inherent compromise in robust design, 
where a slight increase in the expected aleatoric variance is necessary to maintain reliability under epistemic uncertainty.

\begin{table}[h!]
\small
\centering
\caption{Comparison of optimization results for different design points where the target thickness is $240\mathrm{nm}$, and the target probability of failure is 3.0\%.}
\label{tab:optimization_summary}
\begin{tabular}{l c c c c}
\toprule
\textbf{Design} & \textbf{Optimum} & \textbf{Thickness [$\mathrm{nm}$]} & \textbf{Variance [$\mathrm{nm}^2$]} & \textbf{$P_{f}$ [\%]} \\
\midrule
Initial                         & [30.00, 12.50, 80.00] & 243.10 & 1.0631 & 75.35\% \\
DDO                             & [27.84, 12.72, 82.05] & 240.00 & 1.1751 & 1.34\% \\
RBRDO-A                         & [20.00, 12.34, 73.70] & 240.60 & 0.9136 & 2.94\% \\
RBRDO-AE                        & [20.00, 10.49, 74.41] & 240.01 & 0.9626 & 0.71\% \\
\bottomrule
\end{tabular}
\end{table}

\begin{figure}[H]
    \begin{center}
    \includegraphics[width=0.8\textwidth]{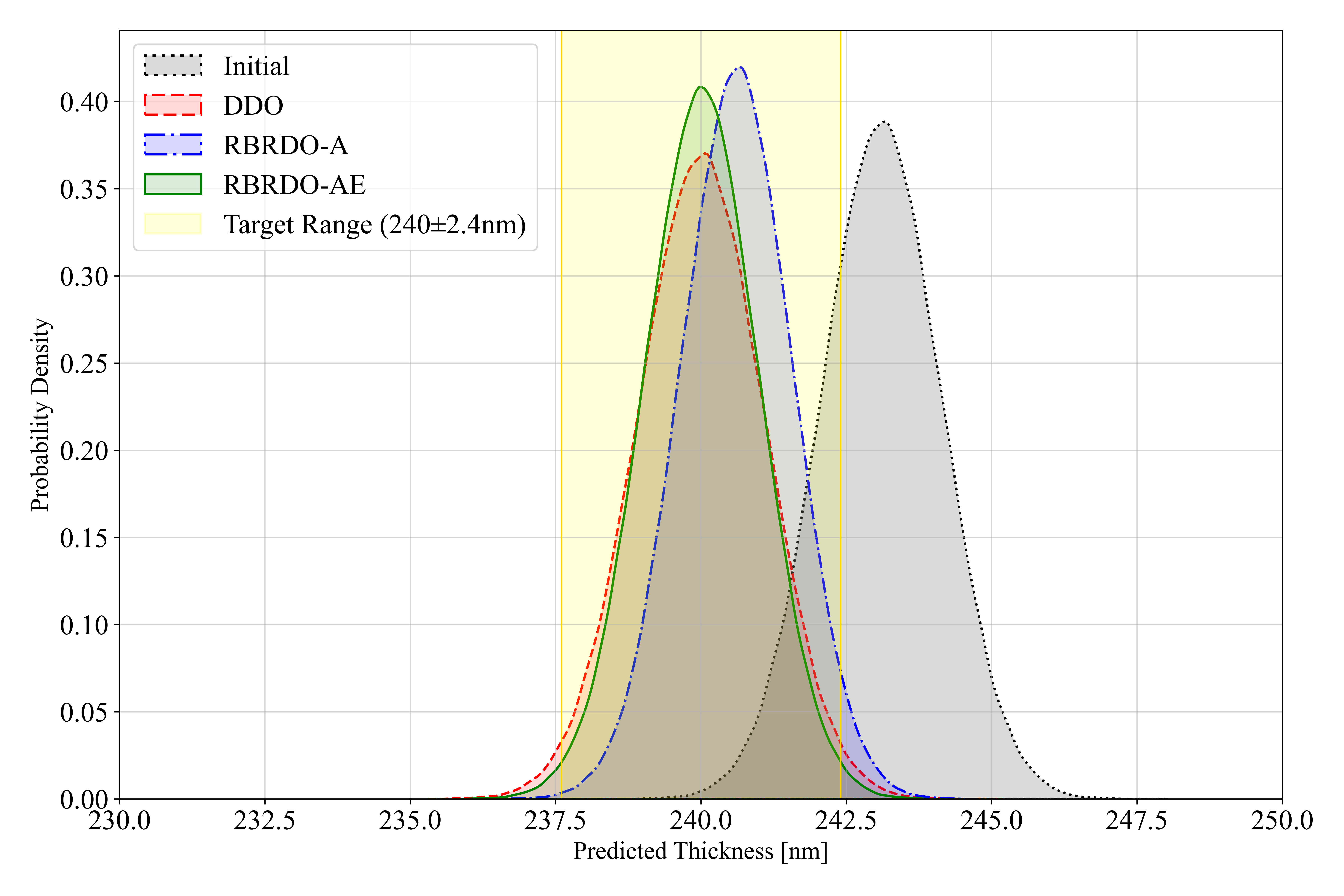}
    \end{center}
    \caption{Predicted probability distributions of thickness at various optimal process parameters.
  }
    \label{fig:comparison_optimum}
\end{figure}

These numerical results are visually demonstrated in \Cref{fig:comparison_optimum}.
The plot shows the predicted thickness distributions from the four design approaches, highlighting again the trade-off between the two RBRDO methods. The optimum of RBRDO accounting for only aleatory uncertainty, denoted as RBRDO-A in \Cref{fig:comparison_optimum}, achieves the narrowest distribution (lowest variance). In contrast, the proposed RBRDO optimum, denoted as RBRDO-AE in \Cref{fig:comparison_optimum}, accepts a slightly higher variance to robustly locate the distribution within the target range, compared to conventional RBRDO.

\begin{figure}[H]
    \centering
    \begin{subfigure}[t]{0.49\textwidth}
        \includegraphics[width=\linewidth]{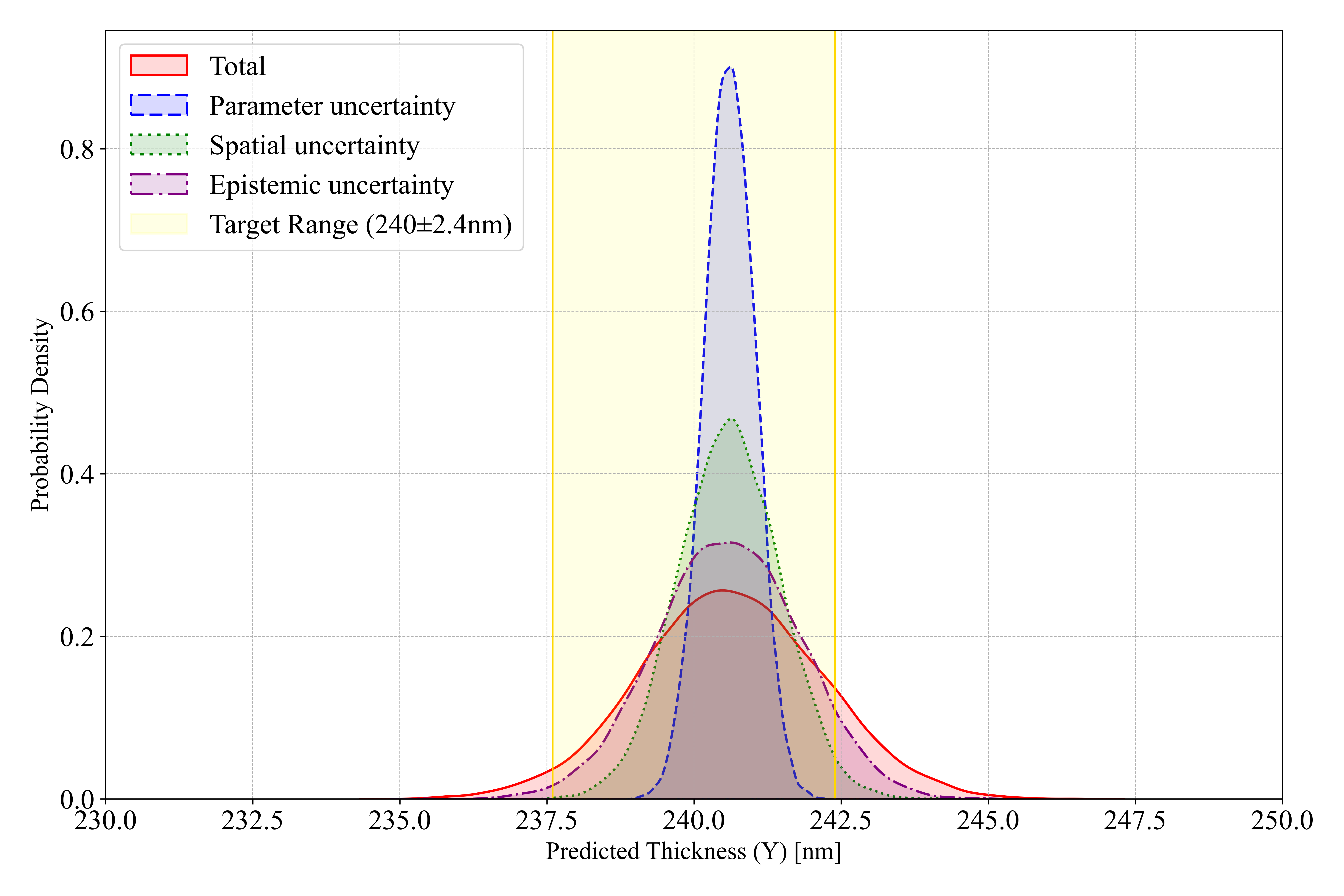}
        \subcaption{}
    \end{subfigure}
    \begin{subfigure}[t]{0.49\textwidth}
        \includegraphics[width=\linewidth]{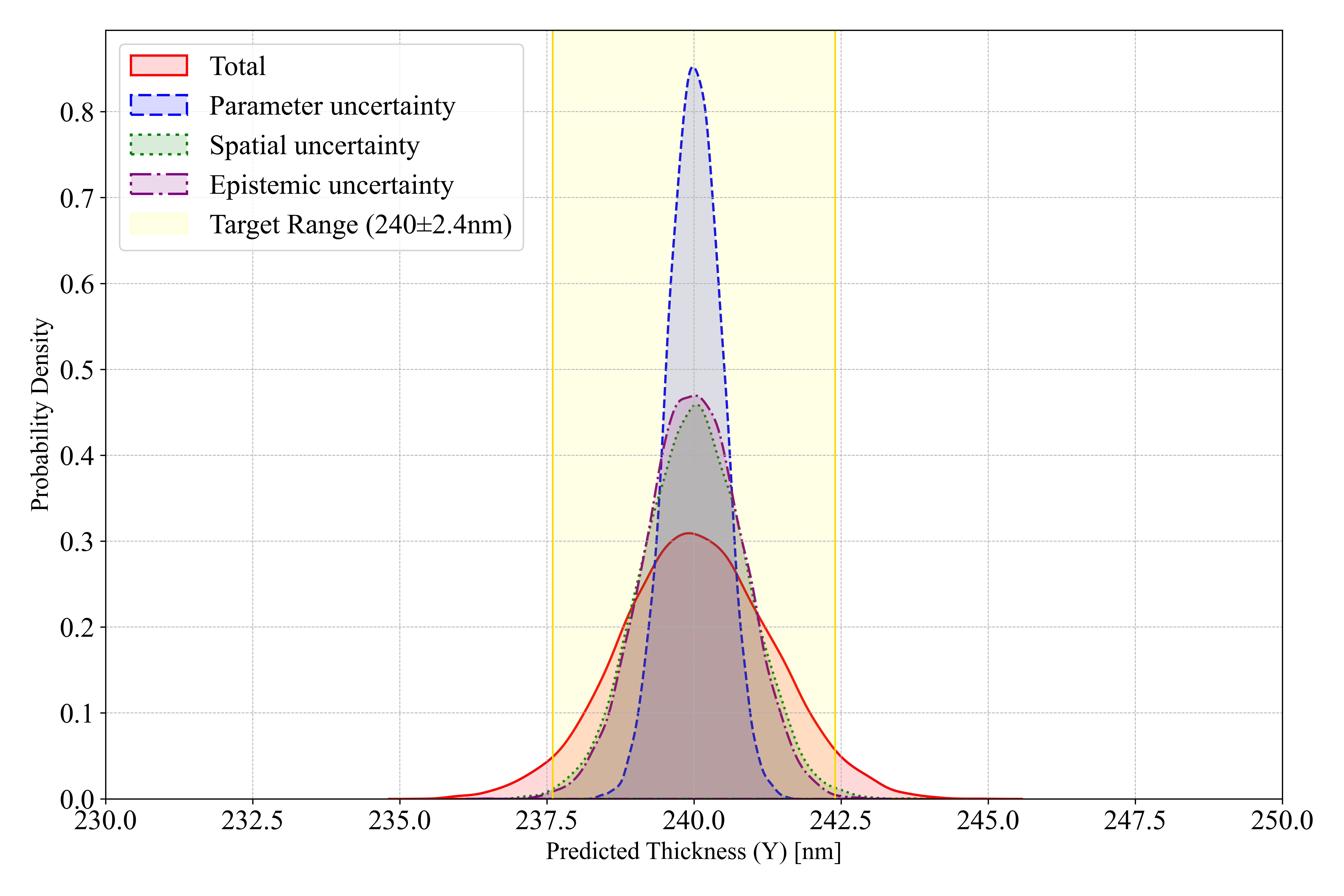}
        \subcaption{}
    \end{subfigure}
    \caption{Uncertainty decomposition of thickness at (a) RBRDO-A optimum and (b) RBRDO-AE optimum.}
    \label{fig:RBRDO_results}
\end{figure}

\begin{figure}[H]
    \begin{center}
    \includegraphics[width=0.8\textwidth]{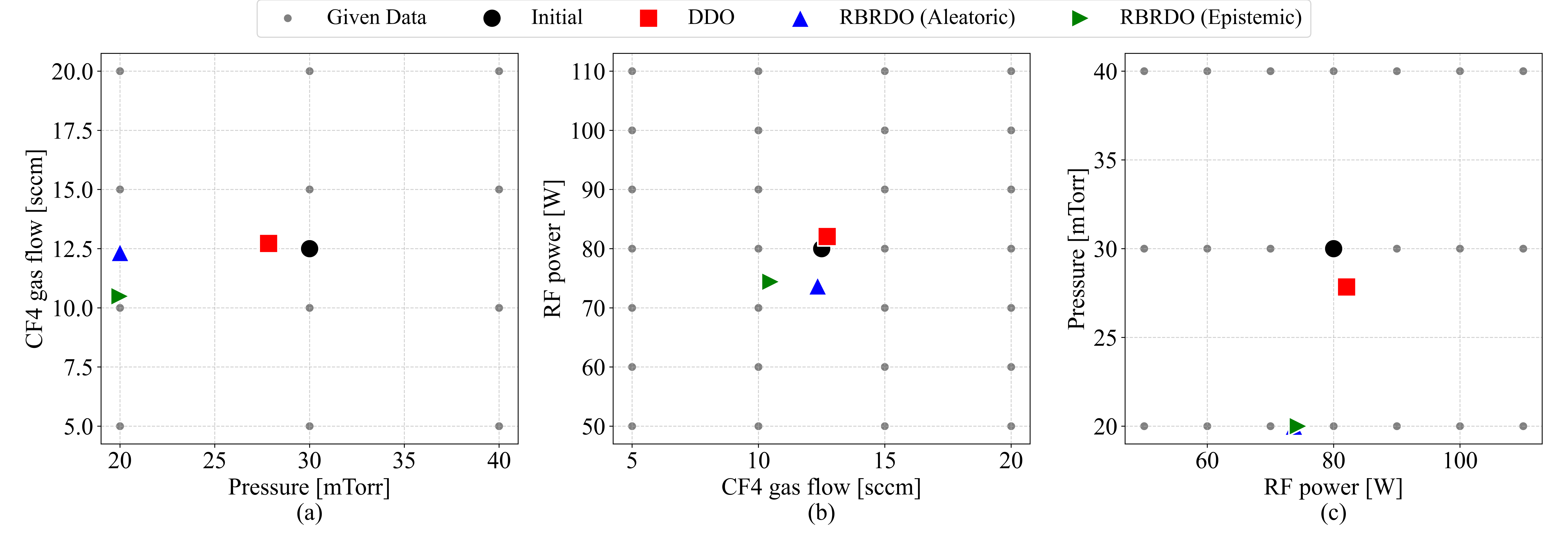}
    \end{center}
    \caption{Visualization of input process parameters for given experimental data and optimum.}
    \label{fig:data_vs_optimum}
\end{figure}

Moreover, \Cref{fig:RBRDO_results} provides further insight by decomposing the uncertainty for the RBRDO-A and RBRDO-AE optimum in \Cref{fig:RBRDO_results}(a) and \Cref{fig:RBRDO_results}(b), respectively. While the relative contributions of the underlying uncertainty sources appear similar, the figures demonstrate that considering epistemic uncertainty leads to a different optimum. This new optimum ensures that the total predictive distribution, which accounts for the combined effect of all uncertainty, robustly satisfies the reliability constraint. This comparison visually confirms that incorporating epistemic uncertainty leads to a more conservative and reliable design choice. In addition, \Cref{fig:data_vs_optimum} shows the location of the optimum relative to the experimental data, which can help guide future experiments to validate the result and reduce the predictive uncertainty.

Lastly, \Cref{tab:optimization_scenarios} summarizes the optimization results for three scenarios under progressively tighter reliability constraints, corresponding to smaller $P_{f,\text{target}}$ values.
In the first two scenarios, the RBRDO-A approach finds the optimum that nominally satisfies the target probability of failure ($P_{f,\text{target}}$), consistently identifying the lowest-variance designs. In contrast, the proposed RBRDO-AE method delivers more conservative solutions, achieving a $P_f$ significantly below the target. This emphasizes a key trade-off to guarantee reliability against the surrogate model's epistemic uncertainty, and thus the proposed method accepts a slightly higher variance, particularly as the target becomes more stringent from Scenario 1 to 2. This distinction becomes critical in Scenario 3, the most demanding case. While the aleatoric-only approach provides a deceptively feasible solution, the proposed framework correctly reports that no feasible solution exists. This outcome demonstrates the method's ability to prevent the selection of designs that appear feasible using the inaccurate hetGP surrogate model but are likely to fail in practice.

\begin{table}[h!]
\small
\centering
\caption{Comparison of optimization results under three scenarios with different target tolerances and probabilities of failure.}
\label{tab:optimization_scenarios}
\begin{tabular}{l c c c c}
\toprule
\textbf{Design} & \textbf{Optimum} & \textbf{Thickness [$\mathrm{nm}$]} & \textbf{Variance [$\mathrm{nm}^2$]} & \textbf{$P_{f}$ [\%]} \\
\cmidrule(r){1-5}
Initial           & [30.00, 12.50, 80.00] & 243.10 & 1.0622 & 31.36 \\
DDO               & [27.84, 12.72, 82.05] & 240.00 & 1.1744 & 0.04  \\
\midrule
\multicolumn{5}{l}{Scenario 1: Target thickness = $240 \pm 3.6$ nm, $P_{f,\text{target}} = 3.0\%$} \\
\cmidrule(r){1-5}
RBRDO-A & [20.00, 12.61, 72.53] & 241.85 & 0.8702 & 3.03  \\
RBRDO-AE & [20.00, 12.51, 73.63] & 240.70 & 0.9117 & 0.10  \\
\midrule
\multicolumn{5}{l}{Scenario 2: Target thickness = $240 \pm 3.6$ nm, $P_{f,\text{target}} = 1.5\%$} \\
\cmidrule(r){1-5}
RBRDO-A & [20.00, 12.64, 72.80] & 241.58 & 0.8800 & 1.53  \\
RBRDO-AE & [20.00, 11.79, 73.86] & 240.39 & 0.9263 & 0.04  \\
\midrule
\multicolumn{5}{l}{Scenario 3: Target thickness = $240 \pm 2.4$ nm, $P_{f,\text{target}} = 1.5\%$} \\
\cmidrule(r){1-5}
RBRDO-A & [20.00, 12.45, 73.97] & 240.33 & 0.9252 & 1.52  \\
RBRDO-AE & \multicolumn{4}{c}{No feasible solution found} \\
\bottomrule
\end{tabular}
\end{table}

\section{Conclusion}

In this study, we proposed the uncertainty quantification and parameter optimization of a plasma etching process using hetGP to effectively separate the aleatory and epistemic uncertainties in complex semiconductor process, and then it enables accurate uncertainty analysis in the process and how to minimize the uncertainty through design optimization. Unlike many existing studies that rely on simulation, our approach was validated using high-fidelity experimental data, providing a more accurate and representative model of the physical process. 

The key contribution of this work is the integration of a hetGP model into an RBRDO of plasma etching process. 
Our method enables the explicit disentanglement and quantification of both aleatory and epistemic uncertainties. The proposed framework utilized a two-stage hetGP model to effectively capture not only the mean thickness but also its input-dependent variance, which corresponds to the spatial variability in wafer. This surrogate model is then employed in variance decomposition to compute the contribution of each uncertainty, reliability analysis for targeting thickness, and RBRDO designed to minimize the variance of the thickness induced by aleatory uncertainty, thereby enhancing process uniformity, while satisfying a probabilistic constraint for a target thickness range. This formulation directly addresses the impact of both inherent process variability (i.e., aleatoric) and model inadequacy due to sparse data (i.e., epistemic), and the results confirmed the effectiveness of the proposed approach.  
Moreover, successfully identified process parameters that significantly improved process robustness by minimizing thickness variation while ensuring reliability. 
This work presents a practical and robust methodology that enables engineers to optimize plasma etching processes, identify sources of variation, and make informed design decisions with a clear understanding of the associated uncertainties.

We also recognize two primary limitations. First, the quantification of spatial variability was based on measurements at only nine discrete locations, which may not fully capture the continuous nature of non-uniformity in the wafer. 
Second, while reliance on experimental data ensures high physical fidelity, this approach is resource-intensive and poses significant practical challenges for large-scale data acquisition and validation.
Future research will extend in two primary directions. 
First, additional experiments will be conducted to cover a wider range of process parameters and to collect more measurements from multiple wafer locations, thereby improving the characterization of spatial variability. 
A multi-fidelity approach will be investigated to combine the experimental results with simulation data, aiming to improve predictive accuracy in regions with sparse experimental measurements.
Second, the proposed method will be extended to capture the time-dependent characteristics of the etching process. This will require the application of methods such as stochastic process modeling and time-variant reliability analysis to predict the process evolution over time, not just the final estimates. Therefore, it aims to establish a comprehensive framework for uncertainty quantification and parameter optimization in semiconductor etching processes. 

\section{Acknowledgement}

This work was supported by the KIMM institutional programs (KN038A, KN037D) and by NST \& KIMM, as well as by the 2025 Daejeon RISE Project (DJR2025-12), funded by the Ministry of Education and Daejeon Metropolitan City, Republic of Korea. J. Kim also acknowledges support from the National Research Foundation of Korea (NRF) grant funded by the Korean government under Grant No. RS-2024-00333943.

\printbibliography

\end{document}